\documentclass[11pt]{article} 

\usepackage[letterpaper, margin=1in]{geometry} 
\usepackage{setspace}
\usepackage{hyperref}

\usepackage{url}
\usepackage{graphicx} 
\usepackage{amssymb,amsmath,amsthm}
\usepackage{thmtools, thm-restate}
\usepackage{booktabs} 
\usepackage[ruled]{algorithm2e} 

\SetAlFnt{\small}
\SetAlCapFnt{\small}
\SetAlCapNameFnt{\small}
\SetAlCapHSkip{0pt}
\IncMargin{-\parindent}

\usepackage{enumitem} 
\usepackage{cases} 
\usepackage{xcolor}
\definecolor{WildStrawberry}{RGB}{255,67,164}
\definecolor{Dousha}{RGB}{47, 136, 67}
\definecolor{BlueBerry}{RGB}{44, 96, 189}

\newtheorem{theorem}{Theorem}
\newtheorem{lemma}{Lemma}
\newtheorem{claim}{Claim}
\newtheorem*{claim*}{Claim}

\newtheorem{corollary}{Corollary}

\newtheorem{proposition}{Proposition}
\newtheorem{remark}{Remark}
\newtheorem*{remark*}{Remark}
\newtheorem*{lemma*}{Lemma}
\newtheorem{fact}{Fact}

\newcommand{\be}{\begin{equation}}
\newcommand{\ee}{\end{equation}}
\newcommand{\beq}{\begin{equation*}}
\newcommand{\eeq}{\end{equation*}}

\newcommand{\R}{\mathbb{R}}

\newcommand{\eps}{\varepsilon}

\newcommand{\AutoAdjust}[3]{\mathchoice{ \left #1 #2  \right #3}{#1 #2 #3}{#1 #2 #3}{#1 #2 #3} }
\newcommand{\Xcomment}[1]{{}}

\newcommand{\InBrackets}[1]{\AutoAdjust{[}{#1}{]}}
\newcommand{\abs}[1]{\AutoAdjust{|}{#1}{|}}%
\newcommand{\Ex}[2][]{\operatorname{\mathbf E}_{#1}\InBrackets{#2}}

\newcommand{\Prx}[2][]{\operatorname{\mathbf{Pr}}_{#1}\InBrackets{#2}}

\newcommand{\eqdef}{\overset{\mathrm{def}}{=\mathrel{\mkern-3mu}=}}
\newcommand{\vect}[1]{\ensuremath{\mathbf{#1}}}

\newcommand{\iprod}[2]{\langle #1, #2 \rangle}
\newcommand\restr[2]{{
  \left.\kern-\nulldelimiterspace 
  #1 
  \vphantom{\big|} 
  \right|_{#2} 
  }}
\def\prob{\Prx}

\DeclareMathOperator{\SO}{SO}

\newcommand{\centi}[1][i]{c_{#1}}
\newcommand{\centers}{\vect{C}}
\newcommand{\loci}[1][i]{x_{#1}}

\newcommand{\loc}{x}
\newcommand{\locs}{\vect{X}}

\newcommand{\Ronedist}[1][R]{d_{_{\text{1,}#1}}}
\newcommand{\twosum}{L_{2}}
\newcommand{\onesum}{L_{1}}
\newcommand{\Ronesum}[1][R]{L_{_{\text{1,}#1}}}
\newcommand{\OPT}{\mathrm{OPT}}      
\newcommand{\Ctwo}{\OPT_2}
\newcommand{\Cone}{\OPT_1}
\newcommand{\CRone}{\OPT_1(R)}
\newcommand{\twoopt}{\twosum(\Ctwo)}
\newcommand{\oneopt}{\onesum(\Cone)}
\newcommand{\Roneopt}{\Ronesum(\CRone)}

\newcommand{\onenorm}[1]{\left\Vert #1\right\Vert_{1}}
\newcommand{\twonorm}[1]{\left\Vert #1\right\Vert_{2}}

\newcommand{\alg}{\texttt{ALG}}

\newcommand{\supp}{\operatorname{supp}}

\title{$L_1$ versus $L_2$ Clustering: When Does the Choice of Norm Matter?}

\author{
Nick Gravin\\
\small Shanghai University of Finance and Economics
\and
Jianhao Jia\\
\small The University of Hong Kong
}

\date{}

\begin{document}
\begin{titlepage}

\maketitle

\begin{abstract}
\normalsize
We study how well approximation guarantees for clustering under one norm translate into guarantees under another. Specifically, we compare $k$-clustering in $\mathbb{R}^d$ under the $L_1$ and $L_2$ objectives, defined as the sums of Manhattan and Euclidean distances, respectively. In the worst case, the elementary relationship between these norms yields a $\sqrt{d}$ approximation bound, which is already tight for $k=2$. We therefore investigate whether stronger guarantees are possible in a natural average-case model, where the dataset is arbitrary but the coordinate system is chosen uniformly at random among all orthonormal bases.

Our main result shows that, when $d=\Omega(\log k)$, with high probability over the random rotation, any $\beta$-approximate $L_1$ clustering, with optimally chosen centers for its clusters, is also an $O(\beta)$-approximation under the $L_2$ objective. We also establish an expected approximation ratio of $O(\beta\min\{\sqrt{d},k^{2/d}\})$, yielding a dimension-independent bound of $O(\beta\sqrt{\log k/\log\log k})$. We obtain analogous guarantees in the reverse direction, from $L_2$ to $L_1$.

We complement these results with lower bounds of $\Omega(\min\{\sqrt{d},k^{1/d}\})$ in both directions, showing that our dimension-independent guarantees have asymptotically optimal dependence on $k$ for constant $\beta$. Our proofs overcome the dependence of the algorithm's output on the random rotation using geometric projection arguments and small Euclidean coresets. In particular, our results provide guarantees for existing clustering algorithms without requiring their redesign.
\end{abstract}


\end{titlepage}


\section{Introduction}
    \label{sec:intro}

Clustering $n$ data points in a $d$-dimensional space is one of the most common problems in 
unsupervised learning. Two standard optimization objectives are the sum of squared Euclidean 
distances ($L_2^2$ objective) for $k$-means clustering and the sum of $L_1$ distances ($L_1$ 
objective) for $k$-median clustering. Which of these objectives is more appropriate in practice 
remains a subject of ongoing debate. Among the most widely used heuristics for these problems 
are Lloyd-type iterative algorithms, often combined with initialization methods such as 
$k$-means++~\cite{Arthur2007kmeans} and its variants.

In practice, the choice between these objectives is often guided by empirical comparisons of the resulting clusterings across different datasets. However, interpreting such comparisons can be challenging, particularly in high-dimensional spaces $\R^d$, where the differences between clusterings are difficult to visualize. Conventional wisdom suggests that $k$-median clustering is more robust to outliers, whereas $k$-means typically enjoys a computational advantage, with faster iterations by a constant factor. Moreover, the $L_2^2$ objective is particularly well suited to data generated from mixtures of spherical Gaussian distributions, whereas the $L_1$ objective may be preferable for data with heavy-tailed or skewed distributions.

The difference between these objectives, and in particular the robustness of the $L_1$ objective to outliers, is easy to understand and visualize in $\R^1$, where the $L_1$ and $L_2$ norms coincide and the only distinction is whether distances are squared. Consider a dataset consisting of $10^4$ points at $-1$, another $10^4$ points at $0$, and just $10$ points at $100$, with $k=2$. The optimal $L_1$ clustering places its centers at $-1$ and $0$, achieving an $L_1$ cost of $10^3$, but an $L_2^2$ cost of $10^5$. In contrast, the optimal $L_2^2$ clustering places its centers at $-1/2$ and $100$, achieving an $L_2^2$ cost of $5\cdot 10^3$, but an $L_1$ cost of $10^4$. Thus, the two objectives lead to different optimal solutions, each performing noticeably worse when evaluated under the other objective. This example illustrates that robustness to outliers can be viewed as a difference in priorities: the $L_1$ objective favors a more refined clustering of populous groups, whereas the $L_2^2$ objective places greater emphasis on accommodating small, remote groups.
    
The preceding example illustrates the effect of squaring distances, even in one dimension, where the $L_1$ and $L_2$ norms coincide. In higher dimensions, however, the norms themselves differ. To isolate the effect of the underlying norm from that of squaring distances, we focus on two unsquared clustering objectives: the sum of Manhattan distances and the sum of Euclidean distances.
\begin{quote}
\emph{How well does a clustering that approximately minimizes the sum of $L_1$ distances perform when evaluated under the $L_2$ objective (without squaring)?}
\end{quote}
We also investigate the converse question:
\begin{quote}
\emph{How well does an approximate solution to the $L_2$ clustering problem perform when evaluated under the $L_1$ objective?}
\end{quote}

The broader question of finding clustering solutions that perform well under multiple objectives has attracted considerable attention in theoretical computer science. In particular, recent work has developed approximation algorithms that simultaneously optimize multiple norms of clustering cost vectors, including general norm objectives for $k$-clustering and all-norm approximation guarantees for correlation clustering~\cite{ChakrabartyS19,DaviesMN24,Cao0Y25}. Related work has also investigated bicriteria and Pareto-optimal approximations for clustering under competing objectives~\cite{ArutyunovaERSSW24,AlamdariS18}. Our work differs from this line of research in two important respects. First, we study the effect of changing the underlying distance metric, rather than the way clustering costs are aggregated. Second, our goal is not to design new algorithms that simultaneously approximate multiple objectives, but rather to understand whether approximation guarantees for one objective translate into guarantees for another. In particular, we investigate how well approximate solutions to the $L_1$ clustering problem perform under the $L_2$ objective, and vice versa, without requiring any modifications to the algorithms producing these solutions.

\paragraph{Baseline Bounds.}
For every vector $v\in\R^d$, the elementary inequalities $\twonorm{v}\leq \onenorm{v}\leq\sqrt{d}\cdot\twonorm{v}$ show that the identity embedding between these metric spaces has distortion at most $\sqrt{d}$. Consequently, any $\alpha$-approximate solution for either clustering objective is automatically an $\alpha\sqrt{d}$-approximation for the other. It is natural to ask whether this dependence on \(d\) can be improved, or even eliminated.
Indeed, a much stronger bound is possible in the simplest case of $k=1$. Here, the optimal $L_1$ solution is the coordinate-wise median of the dataset. For a long time, its best known approximation guarantee under the $L_2$ objective was $\sqrt{d}$, and this dependence on the dimension was even believed to be unavoidable (see, e.g.,~\cite{ZampetakisZ23}). However, recent work~\cite{GravinJia2025} established a dimension-independent approximation ratio of $\sqrt{6\sqrt{3}-8}\approx 1.5467$. Unfortunately, this phenomenon does not extend to $k=2$ in general: there exist simple three-point instances that attain the worst-case bound of $\sqrt{d}$ (see Claim~\ref{cl:mc-without-rotation} and preceding examples).
Nevertheless, these worst-case examples rely on a carefully chosen alignment of the data with the coordinate axes. This motivates investigating whether substantially stronger approximation guarantees can be obtained in a natural average-case setting.

\paragraph{The Random-Rotation Model.}
We consider a natural average-case model in which the dataset is arbitrary, but the coordinate system is chosen at random\footnote{The idea of random rotation of the coordinate system in facility location problems is not new; see, e.g.,~ \cite{meir2019strategyproof,GoelHC23,barakEC26}}. More precisely, we fix an arbitrary set of points in $\R^d$ and choose an orthonormal basis uniformly at random. Equivalently, we apply a uniformly random orthogonal transformation to the dataset. Importantly, such a transformation preserves all Euclidean distances, and hence the $L_2$ clustering objective, while changing the $L_1$ distances between points. Thus, our model allows us to study the effect of the coordinate system on the relationship between the two clustering objectives, without making any distributional assumptions on the dataset itself. We investigate whether, in this setting, approximate solutions to the $L_1$ clustering problem provide substantially better guarantees under the $L_2$ objective than the worst-case $\sqrt{d}$ bound.

\subsection{Our Results}
    \label{sec:results}

We establish approximation guarantees for clustering under the random-rotation model in both directions, showing that randomizing the coordinate system can substantially improve upon the worst-case $\sqrt{d}$ bound. 

\paragraph{From $L_1$ to $L_2$.}
Our main result shows that, when $d=\Omega(\log k)$, with high probability over a uniformly random rotation $R$, any $\beta$-approximate solution to the $L_1$ clustering problem whose centers are coordinate-wise medians of their respective clusters is also an $O(\beta)$-approximation under the $L_2$ objective (Corollary~\ref{cor:l1-l2-constant-probability}).
This result is based on a geometric sufficient condition for such a guarantee to hold (Theorem~\ref{thm:l1-l2-distortion}): it suffices that the pairwise distances between the $k$ optimal $L_2$ centers exhibit constant $L_1$-to-$L_2$ distortion. Combining this geometric insight with the elementary $\sqrt{d}$ bound, we obtain an expected approximation ratio of
\[
O\left(\beta\min\{\sqrt{d},k^{2/d}\}\right)
\]
over the random rotation (Corollary~\ref{cor:l1-l2-expectation}). In particular, maximizing this bound over all dimensions $d$ yields a dimension-independent guarantee of
$
O\left(\beta\sqrt{\frac{\log k}{\log\log k}}\right)
$
for sufficiently large $k$ (Corollary~\ref{cor:l1-l2-k-only}).

\paragraph{From $L_2$ to $L_1$.}
We also investigate the converse direction, which exhibits some notable differences. Already for $k=1$, we show that an optimal $L_2$ center can incur an $L_1$ cost larger than the optimal $L_1$ cost by a factor of $\Theta(d^{1/4})$ in the worst case. This contrasts sharply with the constant-factor guarantee for an optimal $L_1$ center evaluated under the $L_2$ objective~\cite{GravinJia2025}. Nevertheless, under a random rotation, we establish analogous positive results: 
when $d=\Omega(\log k)$, with high probability over a uniformly random rotation $R$, any $\beta$-approximate solution to the $L_2$ clustering problem is also an $O(\beta)$-approximation under the $L_1$ objective (Proposition~\ref{prop:l2-l1-constant-probability}).
Moreover, its expected approximation ratio is
\[
O\left(\beta\min\{\sqrt{d},k^{2/d}\}\right),
\]
and hence at most
$
O\left(\beta\sqrt{\frac{\log k}{\log\log k}}\right)
$
in every dimension (Corollary~\ref{cor:l2-l1-k-only}).

\paragraph{Lower Bounds.}
Finally, we complement these upper bounds with nearly matching lower bounds. For both directions, we construct instances for which the expected approximation ratio over a random rotation is
\[
\Omega\left(\min\{\sqrt{d},k^{1/d}\}\right)
\]
(Theorems~\ref{thm:l1-l2-lower} and~\ref{thm:l2-l1-lower}). Although a gap remains between the exponents of $k$ in our upper and lower bounds, the bounds match asymptotically when maximized over the dimension $d$. In particular, for any constant $\beta$, both yield the same dependence of
$
\Theta\left(\sqrt{\frac{\log k}{\log\log k}}\right)
$
in the worst case over dimensions. Thus, the dependence on $k$ in our dimension-independent guarantee is asymptotically tight for constant-factor approximation algorithms.

\subsection{Technical Overview}  
    \label{sec:technical}

The main challenge in analyzing clustering under a random rotation is that the output of the algorithm, including its clusters and centers, may itself depend on the rotation. Consequently, standard concentration bounds for the $L_1$ norm of a randomly rotated fixed vector cannot be applied directly to the distances between data points and the centers returned by the algorithm. Our main idea is to overcome this difficulty by reducing the analysis to a small collection of directions that are fixed independently of the rotation.

\paragraph{From $L_1$ to $L_2$.}
For this direction, we make the mild assumption that the $L_1$ approximation algorithm chooses an optimal center for each of its clusters, namely, the coordinate-wise median. Equivalently, we may regard the algorithm's output as a partition into clusters, rather than a set of centers (see Remark~\ref{rm:l1-clusters-not-centers}).

The key probabilistic observation is that, for any fixed vector $v\in\R^d$, a uniformly random rotation $R$ satisfies
$
\onenorm{Rv}=\Omega(\sqrt{d})\cdot\twonorm{v}
$
with high probability. Unfortunately, this bound cannot be applied directly to the distances between the algorithm's centers and the data points, since the centers depend on the rotation. Even replacing each center with a uniformly sampled point from its cluster would introduce an undesirable dependence on the number of data points $n$. This suggests compressing the dataset to a small weighted instance, similarly to the use of \emph{coresets} in clustering. However, standard coreset constructions rely on importance sampling tailored to a particular objective, making them less suitable for comparing different norms.

Instead, we exploit a simple geometric reduction: we project every data point onto its nearest center in a fixed optimal $L_2$ clustering, while preserving the partition produced by the $L_1$ approximation algorithm. Using the constant-factor guarantee for coordinate-wise medians under the $L_2$ objective~\cite{GravinJia2025}, we show that the resulting projection introduces only controlled losses in the approximation guarantee. Crucially, the projected instance is supported on at most $k$ fixed points, namely, the optimal $L_2$ centers. Thus, we need to control the distortion of only $O(k^2)$ directions, independently of $n$ and of the algorithm's output. Concentration inequalities for random rotations yield an $O(\beta)$-approximation when $d=\Omega(\log k)$. To obtain an expected approximation guarantee in all dimensions, we bound the probability that the $L_1$ length of a fixed vector becomes unusually small under a random rotation. A union bound over the $O(k^2)$ relevant directions, followed by integration of the resulting tail bound, gives the expected approximation ratio of
$
O\left(\beta\min\{\sqrt{d},k^{2/d}\}\right).
$

\paragraph{From $L_2$ to $L_1$.}
The converse direction requires a different approach. Here, the dependence of the algorithm's output on the rotation can be bypassed using the elementary upper bound of $\beta\sqrt{d}$ times the optimal $L_2$ cost on its $L_1$ cost. The main challenge is therefore to establish a sufficiently strong lower bound on the optimal $L_1$ cost after a random rotation. We again use the idea of compressing the dataset, but this time through a weighted Euclidean coreset, constructed independently of the rotation and containing only $O(k\operatorname{polylog} k)$ distinct points. The coreset preserves the optimal $L_2$ cost up to a constant factor, while, with constant probability, its optimal $L_1$ cost is no larger than a constant multiple of the original optimum. This allows us to reduce the comparison of the two objectives to pairwise distances among the coreset points, whose directions are fixed independently of the rotation. Combining probabilistic estimates for these distances with averaging over the randomness of the coreset construction and the rotation, we obtain analogous approximation guarantees in the reverse direction, albeit without the intuitive geometric characterization available for the first direction.

\paragraph{Lower Bounds.}
Our lower-bound constructions also admit a simple geometric interpretation. We construct instances with only $k+1$ points, whose pairwise directions form a sufficiently fine net of the Euclidean sphere. Consequently, regardless of the rotation, some pair becomes nearly aligned with a coordinate axis, creating a substantial discrepancy between its $L_1$ and $L_2$ lengths. This causes the two objectives to favor different pairs of points to cluster together, yielding the lower bounds of
$
\Omega\left(\min\{\sqrt{d},k^{1/d}\}\right)
$
in both directions.

\subsection{Related Work}
    \label{sec:related}
\paragraph{Simultaneous approximation across norm objectives.}
A long line of work asks whether one clustering can perform well under
many different ways of aggregating client costs. Early results obtained
simultaneous guarantees through resource augmentation, allowing the
algorithm to open more than $k$ centers
~\cite{KumarKleinberg06,GoelMeyerson06,GolovinEtAl08}.
Chakrabarty and Swamy later studied the problem with exactly $k$
centers and gave a constant-factor approximation to the best
simultaneous approximation factor achievable on each instance
~\cite{ChakrabartyS19}. Portfolio formulations relax the requirement of finding
one universal clustering and instead return a small collection of
solutions containing a good choice for every norm
~\cite{GuptaMoondraSingh23,Gupta25}.

Simultaneous norm guarantees have also been studied in other clustering
models. For correlation clustering, Davies, Moseley, and Newman give
one clustering that simultaneously approximates every $\ell_p$ norm
of the per-vertex disagreement costs~\cite{DaviesMN24}. Cao, Li, and Ye
extend this result to all monotone symmetric norms and give efficient
sequential and parallel algorithms~\cite{Cao0Y25}.

\paragraph{Minimum-norm and ordered clustering.}
A related but distinct line of work assumes that a single norm
objective is specified in advance and designs an algorithm for that
objective. Ordered $k$-median, which includes $k$-median, $k$-center,
and $\operatorname{Top}_{\ell}$ objectives as special cases, admits
constant-factor approximation algorithms
~\cite{BSS18,CS18}. Chakrabarty and Swamy developed a general
framework for $k$-clustering under arbitrary monotone symmetric norms
and obtained constant-factor approximation guarantees
~\cite{ChakrabartyS19}. Subsequent work gives parameterized approximation
schemes for broad classes of general-norm clustering objectives
~\cite{AbbasiEtAl23}. More recently, Pulyassary and Swamy use adaptive
sampling to obtain a bicriteria constant-factor approximation for
arbitrary monotone symmetric norms and an $O(\log k)$ approximation
for $\operatorname{Top}_{\ell}$ objectives
~\cite{PulyassarySwamy26}.

The distinction from simultaneous approximation is that the norm is
known to the minimum-norm algorithm, and the returned clustering may
depend on that norm. A simultaneous algorithm must instead use the
same clustering for an entire family of objectives.

\paragraph{Other variants of center-based clustering.}
Approximation algorithms have been developed for many additional
variants of $k$-median and $k$-means, including explainable
clustering, where the partition must be represented by a threshold
decision tree
~\cite{MDRF20,MS21,GPSY23}; fair clustering under representation,
group-cost, or individual-service constraints
~\cite{CKLV17,MV21,MahabadiVakilian20}; clustering with outliers
~\cite{CharikarEtAl01,FriggstadEtAl18}; capacitated and other
constrained clustering problems~\cite{DingXu15}; and clustering in streaming or
limited-memory models~\cite{GuhaEtAl00}. Other work considers
alternative cost functions, including ordered and
$\operatorname{Top}_{\ell}$ objectives
~\cite{BSS18,CS18,ChakrabartyS19}, objectives based on sums of \(p\)-th powers of assignment distances~\cite{CohenAddadKleinMathieu19}, and
Bregman-divergence clustering~\cite{AckermannBloemer09}.     

\section{Preliminaries}
    \label{sec:preliminary}

The instance of the $k$-clustering problem consists of $[n]\eqdef\{1,\ldots,n\}$ data points $\locs\eqdef(\loci)_{i\in[n]}$ located in $\R^d$ space, with $\loci\in\R^d$ for each $i\in[n]$. The clustering algorithm $\alg$ takes as input the points $\locs$ and outputs the set of $k$ centers $\centers=\{\centi[1],\ldots,\centi[k]\}\subset\R^d$. The resulting clustering $(A_1,\ldots,A_k)$ is a partition of $[n]$ data points into $k$ groups with respective centers $(a_1,\ldots,a_k)$, where $\{a_1,\ldots,a_k\}=\centers$.

We study two metrics given by the norms $L_2$ and $L_1$ in $\R^d$. For a 
vector $v=(v_1,\ldots,v_d)\in\R^d$, its $L_2$ norm is 
$\twonorm{v}\eqdef\left(\sum_{j=1}^d v_j^2\right)^{1/2}$ and its $L_1$ 
norm is $\onenorm{v}\eqdef\sum_{j=1}^d\abs{v_j}$. Thus, the respective 
distances in $L_2(\R^d)$ and $L_1(\R^d)$ to the set of centers $\centers$ 
are $d_2(\loc,\centers)\eqdef\min_{\centi\in\centers}\twonorm{x-\centi}$ and $d_1(\loc,\centers)\eqdef\min_{\centi\in\centers}\onenorm{x-\centi}$.
The respective objectives for the $L_2$ norm and the $L_1$ norm are 
$$
\twosum(\centers)
            \eqdef
            \sum_{i\in[n]} d_2(\loci,\centers)
            =
            \sum_{i=1}^n\min_{\centi[j]\in\centers}\twonorm{\loci-\centi[j]},\quad
\onesum(\centers)\eqdef
            \sum_{i\in[n]} d_1(\loci,\centers)
            =
            \sum_{i=1}^n\min_{\centi[j]\in\centers}\onenorm{\loci-\centi[j]}.
$$
We denote the Euclidean optimum solution as $\Ctwo$ and its value as 
$\twoopt$, respectively, the Manhattan optimum as $\Cone$ and its 
value as $\oneopt$. We assume that the algorithm $\alg$ is a $\beta$-approximation to the respective optimum $\twoopt$ for $\alg$ with the $L_2$ objective ($\twosum(\centers)\le\beta\cdot\twoopt$) and $\oneopt$ for $\alg$ with the $L_1$ objective ($\onesum(\centers)\le\beta\cdot\oneopt$), for a constant $\beta\ge 1$. We are interested in comparing (i) $L_2(\alg(\locs))$ against $\twoopt$ for the approximation algorithm $\alg$ with the $L_1$ objective, and (ii)
$L_1(\alg(\locs))$ against $\oneopt$ for the approximation algorithm $\alg$ with the $L_2$ objective.
\paragraph{One Center Problem.}
We begin with the case $k=1$, where the two objective comparisons behave differently. An $L_1$-optimal center (coordinate-wise median) gives a
dimension-independent constant approximation under the $L_2$ objective,
whereas an $L_2$-optimal center (geometric median) may lose a factor of order $d^{1/4}$ under the $L_1$ objective. Specifically,
\begin{fact}[\cite{GravinJia2025}]
\label{fact:one-center}
    For $k=1$, $\twosum(\Cone)\leq\alpha\cdot\twoopt$, where $\alpha=\sqrt{6\sqrt3-8}\approx 1.5467$.
\end{fact}
\begin{remark}
    \label{rm:l1-clusters-not-centers}
    However, if $\alg$ selects a center $c_1$ that is a constant approximation to $L_1(\Cone)$, then $L_2(\{c_1\})$ can be as bad as $\sqrt{d}\cdot L_2(\Ctwo)$. Indeed, it can be easily the case that $L_1(\Cone)=\Omega(\sqrt{d})\cdot L_2(\Cone)$, then by moving $\Cone$ along the first coordinate by $\frac{1}{n}L_1(\Cone)$ to $c_1$ one can still have $2$-approximation to $L_1(\Cone)$, while increasing $L_2$ distance from $L_2(\Cone)$ to $L_2(\{c_1\})=\Omega(1)L_1(\Cone)=\Omega(\sqrt{d})\cdot L_2(\Cone)$.   
\end{remark}
For this reason, we shall assume that the algorithm $\alg$ optimizing the $L_1$ objective produces a locally optimal solution, i.e., the centers $(a_{\ell})_{\ell\in[k]}$ are coordinate-wise medians of the respective clusters $(A_{\ell})_{\ell\in[k]}$. Alternatively, one can view the output of $\alg$ as the clustering $(A_{\ell})_{\ell\in[k]}$ rather than the centers $\centers$ and use the best $L_1$ center per each clusters when evaluating $\alg$ in $L_2$ norm~\footnote{When evaluating the $L_2$ cost, we keep these
coordinate-wise median centers fixed and assign each point
to its Euclidean-nearest center.}.

Next, we evaluate the approximation guarantee of the geometric median under the $L_1$ norm.
\begin{claim}\label{cl:single_cluster_l2_under_l1}
For $k=1$, $\onesum(\Ctwo)= O(d^{1/4})\cdot\oneopt$. This bound is asymptotically tight.
\end{claim}
\begin{proof}[Proof Sketch.] Let $g=\Ctwo$ be the geometric median and $m=\Cone$ be the coordinate-wise median. The triangle inequality reduces the comparison to
bounding the displacement between the two medians: $\onesum(g)\leq\onesum(m)+n\onenorm{g-m}$.
After translating $g$ to the origin and reflecting coordinates so that $m_\ell\geq0$, geometric-median optimality implies that the unit
directional contributions of the points $u_i\eqdef\frac{x_i}{\twonorm{x_i}}$ balance in every coordinate. 
The coordinate-median condition forces a positive directional
contribution of order $\frac{n^2\onenorm{m}}{\twoopt}$, while
the equal negative contribution can be upper bounded by
$O\!\left(\frac{\sqrt d\,\oneopt}{\onenorm{m}}\right)$.
Comparing the two bounds gives $n\onenorm{g-m}
    \leq
    O\!\left(d^{1/4}\sqrt{\twoopt\,\oneopt}\right)$,
and $\twoopt\leq\oneopt$ completes the upper bound.

For the lower bound, we place one point in each of $d$ nearly
coordinate-aligned directions, all sharing a small positive offset in
the remaining coordinates, and add sufficiently many points at the origin. The origin
points keep the geometric median $g$ at $0$, while the repeated positive
coordinates shift the coordinate-wise median $m$ to the common offset.
The Manhattan cost at the geometric median is
$\Theta(d^{5/4})$, whereas the optimal Manhattan cost is
$\Theta(d)$, giving a ratio $\Theta(d^{1/4})$.
The complete proof is deferred to
Appendix~\ref{app:one-center-l2-under-l1}.
\end{proof}

\paragraph{Multiple Centers.} First, consider the elementary distortion bound for the identity embedding $L_1(\R^d)$ in $L_2(\R^d)$,
\begin{align}
    \twosum(\Cone)
    &\leq \oneopt\leq \onesum(\Ctwo)\leq\sqrt d\cdot\twoopt,
    \label{eq:mc-l1-l2-upper}\\
    \onesum(\Ctwo)
    &\leq\sqrt d\cdot\twoopt
    \leq\sqrt d\cdot\twosum(\Cone)\leq\sqrt d\cdot\oneopt.
    \label{eq:mc-l2-l1-upper}
\end{align}
The same norm comparison, together with the $\beta$-approximation guarantee of $\alg$, immediately gives a $\beta\sqrt d$ upper bound in either direction.
With two or more centers, changing the objective can change the groups of
points that are served together, allowing both objective-comparison 
ratios to reach the full $\sqrt d$ factor gap. Consider the following
two instances for $k=2$. Each instance has $k+1=3$ points and two centers,
so the optimal solution is determined by which pair is served together.
The construction uses an axis-aligned pair and a diagonal pair. These
pairs tie in one metric but differ by a factor of $\sqrt d$ in the
other, achieving the full approximation loss.

Let $e_1$ be the first standard basis vector, and let
$\mathbf{1}=(1,\ldots,1)\in\R^d$. The two instances are defined as
follows:
\begin{itemize}
    \item For an $L_1$-optimal solution evaluated under the $L_2$
    objective, place the three points at
    $x_1=0$, $x_2=e_1$, and
    $x_3=-\frac{1}{d}\mathbf{1}$.
    \item For an $L_2$-optimal solution evaluated under the $L_1$
    objective, place the three points at
    $x_1=0$, $x_2=e_1$, and
    $x_3=-\frac{1}{\sqrt d}\mathbf{1}$.
\end{itemize}

\begin{claim}\label{cl:mc-without-rotation}
For the two instances above, there exist tie-breaking rules under
which the $L_1$-optimal solution evaluated under the $L_2$ objective
and the $L_2$-optimal solution evaluated under the $L_1$ objective,
respectively, each attains an approximation ratio of $\sqrt d$.
\end{claim}

\begin{proof}
For the first instance, the pairwise $L_1$ distances are
$\onenorm{\loci[1]-\loci[2]}=\onenorm{\loci[1]-\loci[3]}=1$ and
$\onenorm{\loci[2]-\loci[3]}=2$.
The corresponding Euclidean distances are
$\twonorm{\loci[1]-\loci[2]}=1$,
$\twonorm{\loci[1]-\loci[3]}=\frac1{\sqrt d}$, and
$\twonorm{\loci[2]-\loci[3]}=\sqrt{1+\frac3d}$.
Hence $\oneopt=1$ and $\twoopt=1/\sqrt d$. Fix a tie-breaking rule
that selects $\Cone=\{\loci[1],\loci[3]\}$. Then
$\twosum(\Cone)=1$, giving an approximation ratio of $\sqrt d$.

For the second instance, the pairwise Euclidean distances are
$\twonorm{\loci[1]-\loci[2]}=\twonorm{\loci[1]-\loci[3]}=1$ and
$\twonorm{\loci[2]-\loci[3]}=\sqrt{2+\frac2{\sqrt d}}$.
The corresponding $L_1$ distances are
$\onenorm{\loci[1]-\loci[2]}=1$,
$\onenorm{\loci[1]-\loci[3]}=\sqrt d$, and
$\onenorm{\loci[2]-\loci[3]}=1+\sqrt d$.
Therefore, $\twoopt=\oneopt=1$. Fix a tie-breaking rule that selects
$\Ctwo=\{\loci[1],\loci[2]\}$. Then
$\onesum(\Ctwo)=\sqrt d$, again giving an approximation ratio of
$\sqrt d$.
\end{proof}
For $k>2$, add $k-2$ points sufficiently far from the original
instance and from one another. Each added point is then served by its
own center under both objectives, so the costs and approximation ratio
of the original three-point construction remain unchanged.
\paragraph{Average Case: Random Rotations.}    
The previous bounds of Fact~\ref{fact:one-center}, 
Claim~\ref{cl:single_cluster_l2_under_l1}, and Claim~\ref{cl:mc-without-rotation} hold 
in the worst case over the choice of the orthogonal coordinate system. We are 
interested in the average performance of $\alg$ given 
by a random rotation $R$ of the coordinate system. Formally, we denote the orthogonal 
group of rotations in $\R^d$ as $\SO(d)\eqdef
\left\{R\in\R^{d\times d}: R^\top\cdot R=I_d,\ \det(R)=1 \right\}$, where $I_d$ is the 
$d\times d$ identity matrix, i.e., any $R\in\SO(d)$ preserves $L_2$-distance and inner 
product: $\twonorm{Rv}=\twonorm{v}$ for any $v\in\R^d$, $\iprod{Rv}{Ru}=\iprod{v}{u}$ 
for any $v,u\in\R^d$. Write $\mathbb S^{d-1}\eqdef\{u\in\R^d:\twonorm{u}=1\}$
for the Euclidean unit sphere, equipped with a normalized surface-area measure. 
We assume that random rotation $R\in\SO(d)$ is chosen 
according to the unique Haar measure\footnote{
An intuitive way to understand this measure is to consider the following process. 
Choose the first vector (column) in $R$ sampled uniformly at random from  
$\mathbb{S}^{d-1}$; choose the $i+1$-th vector (column) in $R$ sampled uniformly at 
random from the $\mathbb{S}^{d-i-1}$ unit sphere in $\R^{d-i}$ of vectors orthogonal 
to $\{R_{*1},\ldots,R_{*i}\}$. The sign of its last column should be adjusted to make the determinant positive.} on $\SO(d)$ that essentially picks each row and/or 
column of $R$ uniformly from $\mathbb S^{d-1}$.

The Euclidean distances do not change under any orthogonal rotation. Given a rotation $R$, the Manhattan distance from $\loci\in\R^d$ to $\centers$ is thus $\Ronedist(\loci,\centers)\eqdef\min_{\centi[\ell]\in \centers}\onenorm{R\cdot(\loci-\centi[\ell])}.$ The rotated Manhattan $\Ronesum$ objective ($k$-medians) becomes
\[\Ronesum(\centers)
            \eqdef
            \sum_i\Ronedist(\loci,\centers)
            =
            \sum_{i=1}^n\min_{\centi[j]\in \centers}\onenorm{R(\loci-\centi[j])}. 
\]
We denote the $L_1$ optimum by $\CRone$ and its optimal value by $\Roneopt$. 
Naturally, the output $\centers(R)$ (also the clusters $(A_j(R))_{j\in[k]}$ and their 
centers $(a_j(R))_{j\in[k]}$) of the algorithm $\alg(R,\locs)$ depends on the rotation $R$ when the objective of the algorithm is $\Ronesum$. But even when $\alg(R,\locs)$ optimizes the $L_2$ objective, the output $\centers(R)$ can change with the different coordinate system $R$. We still assume that $\alg$ is a $\beta$-approximation uniformly over $R\in\SO(d)$. That is, when $\alg$ optimizes over $\Ronesum$ we have 
$\Ronesum(\centers(R))\le\beta\cdot\Roneopt$ for $\forall R\in\SO(d)$; when 
$\alg$ optimizes over $L_2$ we have $L_2(\centers(R))\le\beta\cdot\twoopt$.

\paragraph{Probability Estimates for Random Rotations.}
We collect the estimates needed to compare Euclidean and Manhattan
distances after a random rotation. 
For every fixed nonzero $v$, the vector
$Rv/\twonorm{v}$ is uniform on this sphere: its distribution is
unchanged by any fixed rotation. Let $V$ be a uniform unit vector and
define $\lambda_d\eqdef\Ex{\onenorm{V}}$.
The following estimate identifies $\lambda_d$ as the common expected
multiplier and bounds its scale using the Gaussian representation of
the uniform spherical distribution. For completeness, we give the
proofs of all the following propositions in
Appendix~\ref{app:rotation-estimates}.

\begin{proposition}[~\cite{vargaftik2021}, Lemma~9]
\label{lem:rotation-mean}
For every fixed $v\in\R^d$,
    $\Ex[R]{\onenorm{Rv}}
    =
    \lambda_d\twonorm{v}$, where
    $\sqrt{\frac2\pi}\sqrt d
    \leq\lambda_d\leq\sqrt d$.
\end{proposition}

An optimizer chosen after the rotation may favor a segment whose
Manhattan length is unusually small. To control this possibility, we
need a bound on the probability that a fixed segment falls below its
natural $\sqrt d$ scale. A volume comparison between the $L_1$ and
Euclidean unit balls gives the following estimate.

\begin{proposition}[~\cite{Ball1997}, Lecture~1]
\label{lem:rotation-small-ball}
For every fixed nonzero $v\in\R^d$ and every $t\geq1$,
\begin{equation}\label{eq:rotation-small-ball}
    \prob[R]{
        \onenorm{Rv}
        \leq\frac{\sqrt d}{t}\twonorm{v}
    }
    \leq
    \min\left\{1,\left(\frac et\right)^d\right\}.
\end{equation}
\end{proposition}

The compression and sampling arguments used later produce
representative locations before the rotation is drawn. Their pairwise
differences therefore form a fixed family of directions. Although the
selected groups may depend on $R$, a bound that holds for every
direction in this family also controls any nonnegative weighted sum
of their lengths. The following corollary bounds the maximum inverse
distortion over the family.

\begin{corollary}\label{lem:rotation-family}
Let $N\geq1$ and let $\mathcal V$ contain at most $N$ fixed nonzero vectors.
For a nonempty family, define
$\rho(\mathcal V,R)
    \eqdef
    \max\left\{
        1,\;
        \max_{v\in\mathcal V}
        \frac{\sqrt d\,\twonorm{v}}{\onenorm{Rv}}
    \right\}$,
and set $\rho(\mathcal V,R)=1$ when $\mathcal V$ is empty.
Then $1\leq\rho(\mathcal V,R)\leq\sqrt d$, and for every $t\geq1$,
\begin{equation}\label{eq:rotation-family-tail}
    \prob[R]{\rho(\mathcal V,R)>t}
    \leq
    \min\left\{1,N\left(\frac et\right)^d\right\}.
\end{equation}
Consequently, for every $0<\delta<1$, with probability at least
$1-\delta$,
\begin{equation}\label{eq:rotation-family-probability}
    \rho(\mathcal V,R)
    \leq
    \min\left\{\sqrt d,\;e(N/\delta)^{1/d}\right\}.
\end{equation}
\end{corollary}

\begin{proof}
The deterministic bounds follow from norm comparison. For $t\geq1$,
the event $\rho(\mathcal V,R)>t$ requires some $v\in\mathcal V$ to
satisfy
$\onenorm{Rv}<\sqrt d\,\twonorm{v}/t$. A union bound and
Proposition~\ref{lem:rotation-small-ball} give
\eqref{eq:rotation-family-tail}. Taking
$t=e(N/\delta)^{1/d}$ and using the deterministic upper bound
$\sqrt d$ proves~\eqref{eq:rotation-family-probability}.
\end{proof}

The preceding bounds control large inverse distortions. For sharper
constant-factor guarantees in high dimensions, we also use
concentration around the expected length. 

\begin{proposition}[{\cite[Theorem~5.1.4 and Exercise~5.1.12]{Vershynin_2018}}]
\label{lem:rotation-convex-concentration}
There is an absolute constant $\kappa>0$ with the following property.
Let $v_1,\ldots,v_m\in\R^d$ and $b_1,\ldots,b_m\geq0$ be fixed before
$R$, and suppose $D\eqdef\sum_{j=1}^m b_j\twonorm{v_j}>0$.
For every $\eps>0$,
\begin{equation}\label{eq:rotation-convex-concentration}
    \prob[R]{
        \abs{
            \sum_{j=1}^m b_j\onenorm{Rv_j}-\lambda_dD
        }
        >
        \eps\lambda_dD
    }
    \leq
    2\exp(-\kappa\eps^2d).
\end{equation}
\end{proposition}

\section{$L_1$ optimization for the $L_2$ objective}
    \label{sec:l1-under-l2}
Consider an algorithm $\alg$ with a uniform $\beta$-approximation
guarantee under rotation. For a rotation $R$, recall that
$A_1(R),\ldots,A_k(R)$ are the $k$ groups of points returned by the algorithm and $a_j(R)$ is the coordinate-wise median of $A_j(R)$. We analyze the $L_2$ cost of this
solution by projecting the $n$ input points onto the centers of the
$L_2$-optimal solution $\centers=\Ctwo$. For each point $\loci$, let
$c(i)\in \centers$ be its Euclidean-nearest center. Projecting the cluster
$A_j$ onto $\centers$ means replacing each $\loci$ with $i\in A_j$ by
$c(i)$ while keeping the same cluster membership. We denote the resulting projected multiset
$\{c(i):i\in A_j(R)\}$ by $A_j\to \centers$. Its projected Euclidean
and rotated $L_1$ one-center costs are respectively
$$
\twosum(A_j(R)\to \centers)
    \eqdef
    \min_{z\in\R^d}\sum_{i\in A_j(R)}\twonorm{c(i)-z}, 
    \quad\quad
    \Ronesum(A_j(R)\to \centers)
    \eqdef
    \min_{z\in\R^d}\sum_{i\in A_j(R)}\onenorm{R(c(i)-z)}
$$

In the next subsection, we fix the rotation $R$ and prove a
deterministic bound on $\twosum(\alg(R,\locs))$. 
To this end, we first use the one-center bound in Fact~\ref{fact:one-center} to reduce $\twosum(\alg(R,\locs))$ to the sum of
the optimal Euclidean one-center costs of the selected clusters.
Projecting these clusters onto $\centers=\Ctwo$ shows that this sum is at most
$\twoopt+\sum_j\twosum(A_j\to \centers)$. We then compare the $L_2$ and
$L_1$ costs of the projected clusters through the pairwise differences
between centers in $\centers$ exploiting the fact that $\centers$
contains only \(k\) points
(instead of $n$ different locations in $\R^d$ it has only $k$). Finally, the $\beta$-approximation guarantee
bounds their projected $L_1$ cost. Combining these estimates gives the
deterministic comparison.

\subsection{Projection and the fixed-rotation guarantee}
\label{subsec:l1-l2-compression}
Throughout this subsection, we fix the rotation $R$. We write $A_j$
and $a_j$ for $A_j(R)$ and $a_j(R)$, and write $\onesum$ and
$\oneopt$ for $\Ronesum$ and $\Roneopt$. We also write $\twosum(\alg(R,\locs))$ as $\twosum(\alg(\locs))$ and $\onenorm{R(x-y)}$ as
$\onenorm{(x-y)}$. Recall that
$\alpha=\sqrt{6\sqrt3-8}$ is the constant in the one-center bound.

To bound $\twosum(\alg(\locs))$, we first evaluate the returned centers
using the clusters $A_1,\ldots,A_k$. This can only increase the
Euclidean cost, since $\twosum(\alg(\locs))$ assigns each point to its
Euclidean-nearest returned center. Each $a_j$ is a coordinate-wise
median of $A_j$ in the rotated coordinates, so the one-center bound
gives
\begin{equation}\label{eq:l1-l2-group-reduction}
    \twosum(\alg(\locs))
    \leq
    \sum_{j=1}^k\sum_{i\in A_j}\twonorm{\loci-a_j}
    \leq
    \alpha\sum_{j=1}^k
        \min_z\sum_{i\in A_j}\twonorm{\loci-z}.
\end{equation}
To bound $\min_z\sum_{i\in A_j}\twonorm{\loci-z}$, we use the following triangle inequality that holds for every center $z$:
$\sum_{i\in A_j}\twonorm{\loci-z}
    \leq
    \sum_{i\in A_j}\twonorm{\loci-c(i)}
    +
    \sum_{i\in A_j}\twonorm{c(i)-z}$.
Choosing $z$ to minimize the Euclidean cost of the projected group
gives
$\min_z\sum_{i\in A_j}\twonorm{\loci-z}
    \leq
    \sum_{i\in A_j}\twonorm{\loci-c(i)}
    +
    \twosum(A_j\to \centers)$.
Summing over the $k$ groups counts each point once. Since
$\sum_{i=1}^n\twonorm{\loci-c(i)}=\twoopt$,
inequality~\eqref{eq:l1-l2-group-reduction} becomes
\begin{equation}\label{eq:l1-l2-projection-reduction}
    \twosum(\alg(\locs))
    \leq
    \alpha\left(
        \twoopt+\sum_{j=1}^k\twosum(A_j\to \centers)
    \right).
\end{equation}

To bound the remaining sum, we compare the projected $L_2$ and $L_1$
costs. Every pair of projected points either coincides at the same
center of $\centers$ or has a difference in
$\mathcal V=\{c-c':c,c'\in \centers,\ c\neq c'\}$.
Recall that the maximum inverse distortion of these directions under
rotation $R$ is
$\rho(\mathcal V,R)
    =
    \max\left\{
        1,\;
        \max_{v\in\mathcal V}
        \frac{\sqrt d\,\twonorm{v}}{\onenorm{Rv}}
    \right\}$.
Since $R$ is fixed in this subsection, we abbreviate
$\rho(\mathcal V,R)$ by $\rho(\mathcal V)$. Thus every pair of
centers in $\centers$ satisfies
\begin{equation}\label{eq:l1-l2-pair-distortion}
    \twonorm{c(i)-c(\ell)}
    \leq
    \frac{\rho(\mathcal V)}{\sqrt d}
    \onenorm{c(i)-c(\ell)}
    \qquad\text{for all }i,\ell\in[n].
\end{equation}
The following lemma applies this pairwise bound to each projected group $D=A_j\to \centers$.
\begin{lemma}\label{lem:l1-l2-projected-group}
For every nonempty finite multiset $D\subseteq\R^d$, let
$\mathcal V_D\eqdef\{x-y:x,y\in D,\ x\neq y\}$. Then
\[
    \min_z\sum_{x\in D}\twonorm{x-z}
    \leq
    \frac{2\rho(\mathcal V_D)}{\sqrt d}
    \min_z\sum_{x\in D}\onenorm{x-z}.
\]
\end{lemma}
\begin{proof}

Every $y\in D$ is a candidate center for the Euclidean one-center
problem. Averaging the costs of these candidates gives $\min_z\sum_{x\in D}\twonorm{x-z}
    \leq
    \frac1{\abs D}\sum_{x,y\in D}\twonorm{x-y}$.
By definition of $\rho(\mathcal V_D)$,
\[
    \min_z\sum_{x\in D}\twonorm{x-z}
    \leq
    \frac{\rho(\mathcal V_D)}{\sqrt d\,\abs D}
    \sum_{x,y\in D}\onenorm{x-y}.
\]

To bound the $L_1$ pair sum, let $z_D$ minimize
$\sum_{x\in D}\onenorm{x-z}$. For every $x,y\in D$, the triangle
inequality gives
$\onenorm{x-y}\leq
\onenorm{x-z_D}+\onenorm{y-z_D}$.
Summing over ordered pairs, each point-to-center distance appears
$2\abs D$ times. Hence
$\sum_{x,y\in D}\onenorm{x-y}
    \leq
    2\abs D\min_z\sum_{x\in D}\onenorm{x-z}$.
Combining the preceding inequalities gives
\[
    \min_z\sum_{x\in D}\twonorm{x-z}
    \leq
    \frac{2\rho(\mathcal V_D)}{\sqrt d}
    \min_z\sum_{x\in D}\onenorm{x-z}.
\]
\end{proof}
Therefore,
\begin{equation}
\label{eq:l1-l2-projected-group}
    \twosum(A_j\to \centers)
    \leq
    \frac{2\rho(\mathcal V)}{\sqrt d}\,
    \onesum(A_j\to \centers)
    \quad\quad\forall j\in[k]
\end{equation}


Applying~\eqref{eq:l1-l2-projected-group} to
\eqref{eq:l1-l2-projection-reduction}, we obtain
\begin{equation}\label{eq:l1-l2-projected-metric-transfer}
    \twosum(\alg(\locs))
    \leq
    \alpha\left(
        \twoopt+
        \frac{2\rho(\mathcal V)}{\sqrt d}
        \sum_{j=1}^k\onesum(A_j\to \centers)
    \right).
\end{equation}
We conclude by bounding the projected $L_1$ cost using
the centers $a_1,\ldots,a_k$ returned by $\alg$.

\begin{theorem}\label{thm:l1-l2-distortion}
For the fixed rotation $R$,
    $\twosum(\alg(\locs))
    \leq
    \alpha\bigl(1+2(1+\beta)\rho(\mathcal V)\bigr)\twoopt$.
\end{theorem}

\begin{proof}
It remains to bound $\sum_{j=1}^k\onesum(A_j\to \centers)$. For each group $A_j$, use its center $a_j$ as a candidate center for
the projected group. The triangle inequality gives
$\onesum(A_j\to \centers)
    \leq
    \sum_{i\in A_j}\onenorm{c(i)-a_j}
    \leq
    \sum_{i\in A_j}\onenorm{\loci-c(i)}
    +
    \sum_{i\in A_j}\onenorm{\loci-a_j}$.
Summing over the groups and using the $\beta$-approximation guarantee
of $\alg$ gives
\begin{equation}\label{eq:l1-l2-projected-l1-budget}
    \sum_{j=1}^k\onesum(A_j\to \centers)
    \leq
    \sum_{i=1}^n\onenorm{\loci-c(i)}
    +\beta\oneopt.
\end{equation}
The centers in $\centers$ and the assignments $i\mapsto c(i)$ form a feasible solution for $L_1$ objective. Therefore,
$\oneopt\leq\sum_{i=1}^n\onenorm{\loci-c(i)}$, and
\[
    \sum_{j=1}^k\onesum(A_j\to \centers)
    \leq
    (1+\beta)\sum_{i=1}^n\onenorm{\loci-c(i)}
    \leq
    (1+\beta)\sqrt d\,\twoopt.
\]
Substituting this bound into
\eqref{eq:l1-l2-projected-metric-transfer} gives
$\twosum(\alg(\locs))
    \leq
    \alpha\bigl(1+2(1+\beta)\rho(\mathcal V)\bigr)\twoopt$.
\end{proof}

\subsection{Probability and expectation guarantees}
\label{subsec:l1-l2-guarantees}

We use the bound in Theorem~\ref{thm:l1-l2-distortion} from the previous subsection and analyze how the distortion $\rho(\mathcal V,R)$ behaves for a random rotation $R$. 
Note that the upper bound on $L_2(\alg(R,\locs))$ holds for every $R$, since $\alg$ has a uniform
$\beta$-approximation guarantee under each rotation $R$.
The family $\mathcal V$ is determined by the centers of $\Ctwo$
and is therefore fixed before the rotation is drawn. Thus the only
random quantity in the fixed-rotation bound is
$\rho(\mathcal V,R)$. We first use concentration to obtain a constant
approximation with high probability, and then integrate the distortion
tail bound to bound the expected cost.

\begin{corollary}\label{cor:l1-l2-constant-probability}
There are absolute constants $C_0,c_0>0$ such that, for $k\geq1$ and
$d\geq C_0(1+\log k)$, with probability at least $1-e^{-c_0d}$,
    $\twosum(\alg(R,\locs))
    \leq
    9\alpha\beta\twoopt$.
\end{corollary}

\begin{proof}
Apply Proposition~\ref{lem:rotation-convex-concentration} with
$m=1$, $b_1=1$, and $\eps=1/3$ to an arbitrary fixed nonzero vector
$v\in\mathcal V$. For every such
$v$,
$\prob[R]{\onenorm{Rv}<\frac23\lambda_d\twonorm{v}}
\leq2e^{-\kappa d/9}$.
Since $\mathcal V$ contains at most $k^2$ vectors, a union bound shows
that, except with probability at most $2k^2e^{-\kappa d/9}$, $\onenorm{Rv}
    \geq
    \frac23\lambda_d\twonorm{v}$ holds for every $v\in\mathcal V$.
On this event,
$\rho(\mathcal V,R)\leq
\frac{3\sqrt d}{2\lambda_d}
\leq\frac32\sqrt{\frac{\pi}{2}}$,
where the second inequality uses
$\lambda_d\geq\sqrt{2/\pi}\sqrt d$.

Substituting this bound into
Theorem~\ref{thm:l1-l2-distortion} gives
\begin{multline*}
    \twosum(\alg(R,X))\leq
    \alpha\bigl(1+2(1+\beta)\rho(\mathcal V,R)\bigr)\twoopt\\
    \leq
    \alpha\left(
        1+3(1+\beta)\sqrt{\frac{\pi}{2}}
    \right)\twoopt
    \leq
    9\alpha\beta\twoopt,
\end{multline*}
where the last inequality uses $\beta\geq1$.
It remains to bound the failure probability. Since
$\log(2k^2)\leq2(1+\log k)$,
$2k^2e^{-\kappa d/9}\leq\exp\left(2(1+\log k)-\frac{\kappa d}{9}\right)$.
Taking $C_0=36/\kappa$ and $c_0=\kappa/18$, the assumption
$d\geq C_0(1+\log k)$ implies
$2(1+\log k)\leq\kappa d/18$. Hence the failure probability is at
most $e^{-c_0d}$.
\end{proof}

We then obtain the following expectation upper bound from
the tail bound for $\rho(\mathcal V,R)$.

\begin{corollary}\label{cor:l1-l2-expectation}
There is an absolute constant $C>0$ such that, for every $k\geq1$
and $d\geq2$,
\begin{equation}\label{eq:l1-l2-expectation}
    \frac{\Ex[R]{\twosum(\alg(R,\locs))}}{\twoopt}
    \leq
    C\beta\min\{\sqrt d,k^{2/d}\}.
\end{equation}
\end{corollary}

\begin{proof}
Corollary~\ref{lem:rotation-family} gives
$\prob[R]{\rho(\mathcal V,R)>t}
\leq\min\{1,k^2(e/t)^d\}$ for every $t\geq1$.
Using the tail-integral formula and splitting the integral at
$ek^{2/d}$, we obtain
\begin{align*}
    \Ex[R]{\rho(\mathcal V,R)}=\int_0^\infty
        \prob[R]{\rho(\mathcal V,R)>t}dt\leq
        ek^{2/d}
        +k^2e^d\int_{ek^{2/d}}^\infty t^{-d}\,dt=ek^{2/d}\left(1+\frac1{d-1}\right)
    \leq2ek^{2/d}.
\end{align*}
Taking expectations in
Theorem~\ref{thm:l1-l2-distortion} gives
\begin{multline*}
    \frac{\Ex[R]{\twosum(\alg(R,\locs))}}{\twoopt}\leq
    \alpha\left(
        1+2(1+\beta)\Ex[R]{\rho(\mathcal V,R)}
    \right)\leq
    \alpha\left(
        1+4e(1+\beta)k^{2/d}
    \right)
    \leq
    C\beta k^{2/d}.
\end{multline*}

We also have a pointwise bound of order $\beta\sqrt d$. Evaluating the
returned centers using their clusters $A_1(R),\ldots,A_k(R)$ and
applying norm comparison gives
\begin{multline*}
    \twosum(\alg(R,\locs))\leq
    \sum_{j=1}^k\sum_{i\in A_j(R)}
        \twonorm{\loci-a_j(R)}\leq
    \sum_{j=1}^k\sum_{i\in A_j(R)}
        \onenorm{R(\loci-a_j(R))}\\ \leq\beta\Roneopt\leq
    \beta\sqrt d\,\twoopt.
\end{multline*}
Taking the smaller of the two bounds proves
\eqref{eq:l1-l2-expectation}.
\end{proof}

Optimizing the preceding bound over the dimension gives a guarantee
that depends only on $k$.

\begin{corollary}\label{cor:l1-l2-k-only}
For $k\geq3$, uniformly over all $d\geq2$,
    $\frac{\Ex[R]{\twosum(\alg(R,\locs))}}{\twoopt}=
    O\left(\beta\sqrt{\frac{\log k}{\log\log k}}
    \right).$
\end{corollary}

\begin{proof}
By Corollary~\ref{cor:l1-l2-expectation}, it is enough to maximize
$\min\{\sqrt d,k^{2/d}\}$ over $d$. The first term increases with
$d$, while the second decreases. They are equal when
$\sqrt d=k^{2/d}$, or equivalently when
$d\log d=4\log k$. The solution satisfies
$d=\Theta(\log k/\log\log k)$, and hence
$\sqrt d=\Theta(\sqrt{\log k/\log\log k})$.
Away from this intersection, the smaller of the two terms can only
decrease, which concludes the proof.
\end{proof}

\subsection{A lower-bound construction}
\label{subsec:l1-l2-lower}

Our upper bounds apply to any $\beta$-approximate $L_1$
solution.
We now give a lower bound on  the approximation ratio even when the returned solution is an exact $L_1$ optimum for every rotation.

\begin{theorem}\label{thm:l1-l2-lower}
For every $k,d\geq2$, there is an instance with $n=k+1$ points such
that every exact rotated $L_1$ optimum has Euclidean approximation
ratio $\Omega(\min\{\sqrt d,k^{1/d}\})$ with probability at least
$1-(e/4)^d$ over $R$. On the same instance, the expected Euclidean
approximation ratio is also
$\Omega(\min\{\sqrt d,k^{1/d}\})$.
\end{theorem}

To construct such an instance, we use an $\eps$-net of the Euclidean
unit sphere. A set $U\subseteq\mathbb S^{d-1}$ is an $\eps$-net if,
for every $v\in\mathbb S^{d-1}$, there is some $u\in U$ satisfying
$\twonorm{u-v}\leq\eps$. The following standard proposition bounds
the size of such a net and therefore the number of candidate
directions needed in our construction.

\begin{proposition}[{\cite[Corollary~4.2.13]{Vershynin_2018}}]
\label{lem:l1-l2-sphere-net}
For every $0<\eps\leq1$, the Euclidean unit sphere
$\mathbb S^{d-1}$ has an $\eps$-net $U$ satisfying
$\abs U\leq(1+2/\eps)^d\leq(3/\eps)^d$.
\end{proposition}

We next define the instances parameterized by $T$ and $\eps$.

\paragraph{Instance.}
Let $e_1=(1,0,\ldots,0)$, and let
$U=\{u_1,\ldots,u_s\}$ be an $\eps$-net satisfying $2s\leq k-1$.
For $r=1,\ldots,k-1-s$, define
$b_r\eqdef1000r\sqrt d\,e_1$. The instance consists of the following
$k+1$ points:
\begin{itemize}
    \item the special pair $-e_1/2$ and $e_1/2$, whose Euclidean
    distance is $1$;
    \item for every $r\in[s]$, the candidate pair
    $b_r-\frac T2u_r$ and $b_r+\frac T2u_r$, whose Euclidean distance
    is $T$;
    \item one singleton point at each location
    $b_{s+1},\ldots,b_{k-1-s}$.
\end{itemize}
The instance contains
$2+2s+(k-1-2s)=k+1$ points. The sites $b_r$ are sufficiently far
apart that a pair containing points from different sites cannot be a
closest pair in either metric. 
Note that for an instance consisting of \(k+1\) points and allowing \(k\) unrestricted centers, the optimal clustering cost under any metric equals the minimum pairwise distance between distinct input points.

The following claim chooses $T$ and $\eps$ and verifies the required
bounds for this construction.
\begin{claim*}\label{cl:l1-l2-lower-construction}
There are choices of $T$ and $\eps$ such that the rotated $L_1$
optimum $\CRone$ satisfies
\[
    \prob[R]{
        \twosum(\CRone)
        \geq
        \frac{\min\{\sqrt d,k^{1/d}\}}{50}
    }
    \geq
    1-\left(\frac e4\right)^d
    \quad\text{and}\quad
    \Ex[R]{\twosum(\CRone)}
    \geq
    \frac{\min\{\sqrt d,k^{1/d}\}}{50}.
\]
\end{claim*}

\begin{proof}
Set
$T=\frac1{25}\min\{\sqrt d,k^{1/d}\}$.
If $T\leq2$, take any instance with $k+1$ distinct points and
positive Euclidean optimal cost, and scale the instance so that
$\twoopt=1$. Every Euclidean approximation ratio is at least one, and
$\min\{\sqrt d,k^{1/d}\}\leq50$. Hence, for every rotation,
\[
    \twosum(\CRone)
    \geq
    \twoopt
    =
    1
    \geq
    \frac{\min\{\sqrt d,k^{1/d}\}}{50},
\]
so both conclusions follow. We therefore assume $T>2$ and set
$\eps=1/(5T)$.

We first verify that the construction uses at most $k+1$ points.
Since $T>2$, Proposition~\ref{lem:l1-l2-sphere-net} gives an
$\eps$-net of size
\begin{equation}\label{eq:l1-l2-lower-direction-budget}
    s
    \leq
    (1+10T)^d
    \leq
    \left(\frac{21T}{2}\right)^d
    \leq
    k\left(\frac{21}{50}\right)^d
    \leq
    \frac{k}{4}
    \leq
    \frac{k-1}{2}.
\end{equation}
Thus $2s\leq k-1$, as required.

The special pair is the unique Euclidean closest pair. Since the
instance has $k+1$ points and $k$ centers, the Euclidean optimum serves
the special pair together and every other point separately. Hence
$\twoopt=1$.

We next show that every rotation makes some candidate pair short in
the $L_1$ metric. By the definition of the net, for every $R$ there
is some $u_r\in U$ such that
$\twonorm{u_r-R^{-1}e_1}\leq1/(5T)$. Since $R$ preserves Euclidean
distance, $\twonorm{Ru_r-e_1}\leq1/(5T)$. The corresponding candidate
pair therefore has rotated $L_1$ distance
\begin{multline}
    T\onenorm{Ru_r}
    \leq
    T\onenorm{e_1}
    +
    T\onenorm{Ru_r-e_1}
    \leq
    T+T\sqrt d\,\twonorm{Ru_r-e_1}\\
    \leq
    T+\frac{\sqrt d}{5}
    \leq
    \left(\frac1{25}+\frac15\right)\sqrt d
    =
    \frac6{25}\sqrt d
    <
    \frac{\sqrt d}{4}.
    \label{eq:l1-l2-lower-short-candidate}
\end{multline}

Now consider the event
$\mathcal A\eqdef\{\onenorm{Re_1}\geq\sqrt d/4\}$.
On this event, the special pair has rotated $L_1$ distance at least
$\sqrt d/4$, while~\eqref{eq:l1-l2-lower-short-candidate} gives a
candidate pair with strictly smaller $L_1$ distance. Because the
different sites are far apart, every exact rotated $L_1$ optimum
serves a candidate pair $x,y$ together.

Let $z$ be the center serving $x$ and $y$. Both endpoints are within
Euclidean distance $T$ of $z$. All other returned centers are at
different, remote sites, so both endpoints still use $z$ under
Euclidean evaluation. Hence
$\twonorm{x-z}+\twonorm{y-z}\geq
    \twonorm{x-y}=
    T$,
which gives $\twosum(\CRone)\geq T$.

Finally, Proposition~\ref{lem:rotation-small-ball}, applied to $e_1$
with $t=4$, gives
\[
    \prob[R]{\mathcal A}
    =
    1-\prob[R]{\onenorm{Re_1}<\sqrt d/4}
    \geq
    1-\left(\frac e4\right)^d.
\]
On $\mathcal A$,
\[
    \twosum(\CRone)
    \geq
    T
    =
    \frac1{25}\min\{\sqrt d,k^{1/d}\}
    \geq
    \frac1{50}\min\{\sqrt d,k^{1/d}\}.
\]
Therefore,
\[
    \prob[R]{
        \twosum(\CRone)
        \geq
        \frac{\min\{\sqrt d,k^{1/d}\}}{50}
    }
    \geq
    1-\left(\frac e4\right)^d.
\]

Since $d\geq2$ implies $1-(e/4)^2>1/2$,
\[
    \Ex[R]{\twosum(\CRone)}
    \geq
    T\prob[R]{\mathcal A}
    \geq
    \frac T2
    =
    \frac1{50}\min\{\sqrt d,k^{1/d}\}.
\]
\end{proof}
Further maximizing over $d$ yields the following asymptotically tight bound on $k$.
\begin{corollary}\label{cor:l1-l2-tight-k-dependence}
For every $k\geq 3$, the worst-case expected approximation ratio of an
exact rotated $L_1$ optimum under the $L_2$ objective is
$\Theta\left(
        \sqrt{\frac{\log k}{\log\log k}}
    \right)$.
\end{corollary}

\begin{proof}
Corollary~\ref{cor:l1-l2-k-only}, applied
with $\beta=1$, gives the upper bound.
For the lower bound, Theorem~\ref{thm:l1-l2-lower} gives, for every
$d\geq2$, an instance whose expected approximation ratio is
$\Omega(\min\{\sqrt d,k^{1/d}\})$. Choose $d$ near the solution of
$d\log d=2\log k$, for which
$d=\Theta(\log k/\log\log k)$ and
$\sqrt d=k^{1/d}$ up to absolute constant factors. Thus,
\[
    \sup_{d\geq2}\min\{\sqrt d,k^{1/d}\}
    =
    \Omega\left(
        \sqrt{\frac{\log k}{\log\log k}}
    \right).
\]
Combining the two bounds proves the claim.
\end{proof}

\section{$L_2$ optimization for the $L_1$ objective}
    \label{sec:l2-under-l1}

In this section, consider an algorithm $\alg$ with a uniform
$\beta$-approximation guarantee for the $L_2$ objective:
$\twosum(\alg(R,\locs))\leq\beta\twoopt$ for every rotation $R$.
We evaluate the same returned centers under the rotated $L_1$
objective.
Our goal is to bound
$L_{1,R}(\alg(R,\locs))/\Roneopt$. Norm comparison
and the $\beta$-approximation guarantee immediately give:
\begin{equation}\label{eq:l2-l1-numerator}
    \Ronesum(\alg(R, \locs))
    \leq
    \sqrt d\,\twosum(\alg(R, \locs))
    \leq
    \beta\sqrt d\,\twoopt.
\end{equation}

Thus, the remaining task is to lower-bound $\Roneopt$ using $\twoopt$. To compare these two costs, we construct a small weighted
coreset $S$ 
through the standard sensitivity-sampling procedure \textbf{for the $L_2(\R^d)$ metric}
analyzed in~\cite[Algorithm~1 and Appendix~G, Theorem~7]{BCCPS24}. We use $L_{1,R}(\centers;S)$ and $L_{2}(\centers;S)$ to denote the $L_1$ cost and $L_2$ cost on the coreset $S$ given centers $\centers$ and rotation $R$. Write the optimal $L_1$ and $L_2$ centers for the coreset $S$ as $\OPT_1(R,S)$ and $\OPT_2(S)$. The weighted sample $S$ has the following properties:
\begin{enumerate}
    \item The sampling procedure uses at most
    $M\leq C_s k\log^b(ek)$ different locations sampled from $\locs$ for some absolute constants $C_s\geq1$ and $b>0$.

    \item The sampling probabilities depend only on the Euclidean
    instance. We draw $S$ independently of the rotation $R$.

    \item With probability at least $2/3$,
    \begin{equation}\label{eq:l2-l1-euclidean-coreset}
        \frac34\twosum(\centers)
        \leq
        \twosum(\centers;S)
        \leq
        \frac54\twosum(\centers)
        \qquad
        \text{for every }\abs{\centers}=k.
    \end{equation}

    \item For every fixed rotation $R$,
    \begin{equation}\label{eq:l2-l1-sampled-optimum-expectation}
        \Ex[S]{L_{1,R}(\OPT_1(R,S);S)}
        \leq
        \Roneopt.
    \end{equation}
\end{enumerate}

The first and third properties follow from
\cite[Algorithm~1 and Appendix~G, Theorem~7]{BCCPS24}, applied with
$\eps=1/4$. The second property holds because the sampling probabilities depend
only on Euclidean distances, and the sampling randomness is chosen
independently of $R$. The last property is a well-known feature of many coreset constructions and follows from the unbiased inverse-probability weights. We defer the proof to Appendix~\ref{app:coreset_property}. 

In the next subsection, we use the properties of the sensitivity sampling to show that $\Roneopt$ can be bounded by $\twoopt$ through $L_{1,R}(\OPT_1(R,S);S)$ and $L_2(\OPT_2(S);S)$.



\subsection{Sampling and the fixed-rotation guarantee}
\label{subsec:l2-l1-sampling}

Throughout this subsection, we fix the rotation $R$ and omit $R$
from the notation. We write $\alg(X)$, $L_1(C)$,
$L_1(C;S)$, $\OPT_1$, and $\OPT_1(S)$ for
$\alg(R,X)$, $L_{1,R}(C)$, $L_{1,R}(C;S)$,
$\OPT_1(R)$, and $\OPT_1(R,S)$, respectively. Under this convention, $\onenorm{x-y}$ denotes
$\onenorm{R(x-y)}$ throughout the subsection.

Our remaining task is to lower-bound $\oneopt$. Inequality~\eqref{eq:l2-l1-sampled-optimum-expectation} and Markov's inequality give
\begin{equation}\label{eq:l2-l1-sampled-l1}
    \prob[S]{L_1(\OPT_1(S);S)\leq2\oneopt}
    \geq
    \frac12.
\end{equation}

On this event, a lower bound on $L_1(\OPT_1(S);S)$ gives a lower bound on
$\oneopt$. We obtain such a bound by comparing the two objectives on
the same $S$.

For a realized sample $S$, let
$\mathcal V_S\eqdef
\{x-y:x,y\in\supp S,\ x\neq y\}$.
Recall that the maximum inverse distortion of these directions under
the rotation is $\rho(\mathcal V_S,R)
    \eqdef
    \max\left\{
        1,\;
        \max_{v\in\mathcal V_S}
        \frac{\sqrt d\,\twonorm{v}}{\onenorm{Rv}}
    \right\}$.
Since $R$ is fixed throughout this subsection, we write
$\rho(\mathcal V_S)$ for $\rho(\mathcal V_S,R)$. The proof of Lemma~\ref{lem:l1-l2-projected-group} extends directly
to weighted samples. Applying this weighted comparison to every cluster of the sampled
$L_1$-optimal partition and summing gives a feasible Euclidean
$k$-clustering of $S$. Its cost is at least
$L_2(\OPT_2(S);S)$, and therefore
\begin{equation}\label{eq:l2-l1-sample-comparison}
    L_1(\OPT_1(S);S)
    \geq\frac{\sqrt d}{2\rho(\mathcal V_S)}
    L_2(\OPT_2(S);S).
\end{equation}




We can now combine~\eqref{eq:l2-l1-numerator}, the sampling properties, and~\eqref{eq:l2-l1-sample-comparison} and get the following claim for a fixed coordinate system.

\begin{claim}\label{cl:l2-l1-distortion}
With probability at least $1/6$ over the
sample $S$,
\begin{equation}\label{eq:l2-l1-fixed-rotation}
    \frac{\onesum(\alg(\locs))}{\oneopt}
    \leq
    \frac{16}{3}\beta\rho(\mathcal V_S).
\end{equation}
\end{claim}

\begin{proof}
By~\eqref{eq:l2-l1-sampled-l1}, with probability at least $1/2$,
the sampled $L_1$ optimum satisfies
$L_1(\OPT_1(S);S)\leq2\oneopt$. To use
Equation~\eqref{eq:l2-l1-sample-comparison} to lower-bound $\oneopt$, we
also need the sample to retain a constant fraction of the Euclidean
optimal cost. The property of sampling gives
\begin{equation}\label{eq:l2-l1-sampled-l2}
    \prob[S]{
        L_2(\OPT_2(S);S)\geq\frac34\twoopt
    }
    \geq
    \frac23.
\end{equation}
Therefore, the two inequalities
$L_1(\OPT_1(S);S)\leq2\oneopt$ and
$L_2(\OPT_2(S);S)\geq\frac34\twoopt$ hold simultaneously with probability
at least
$\frac12+\frac23-1=\frac16$.
On this event~\eqref{eq:l2-l1-sample-comparison} gives
\[
    2\oneopt
    \geq
    L_1(\OPT_1(S);S)
    \geq
    \frac{\sqrt d}{2\rho(\mathcal V_S)}
    L_2(\OPT_2(S);S)
    \geq
    \frac{3\sqrt d}{8\rho(\mathcal V_S)}
    \twoopt.
\]
Thus
$\oneopt\geq
\frac{3\sqrt d}{16\rho(\mathcal V_S)}\twoopt$.
Combining this lower bound with~\eqref{eq:l2-l1-numerator}, we obtain
\[
    \frac{\onesum(\alg(\locs))}{\oneopt}
    \leq
    \frac{
        \beta\sqrt d\,\twoopt
    }{
        3\sqrt d\,\twoopt/
        (16\rho(\mathcal V_S))
    }
    =
    \frac{16}{3}\beta\rho(\mathcal V_S).
\]
\end{proof}

Although~\eqref{eq:l2-l1-fixed-rotation} still depends on the sampled
$S$, every sample has at most
$M\leq C_s k\log^b(ek)$ support points. Hence
$\mathcal V_S$ contains at most $M^2$ directions. 
We next analyze how the randomness of $R$ affects the bound in~\eqref{eq:l2-l1-fixed-rotation}.

\subsection{Probability and expectation guarantees}
\label{subsec:l2-l1-guarantees}

We use the bound in Claim~\ref{cl:l2-l1-distortion} from the previous subsection and analyze how the distortion $\rho(\mathcal V_S,R)$ behaves for a random rotation $R$. For every fixed rotation,
Claim~\ref{cl:l2-l1-distortion} shows that at least a $1/6$
fraction of the samples satisfy
\[
    \frac{\Ronesum(\alg(R,\locs))}{\Roneopt}
    \leq
    \frac{16}{3}\beta\rho(\mathcal V_S,R).
\]
For every fixed sample $S$, the family $\mathcal V_S$ is independent
of $R$ and contains at most $M^2$ vectors. We combine these two facts
to control the approximation ratio over a random rotation.

\begin{proposition}\label{prop:l2-l1-constant-probability}
There are absolute constants $C_0,c_0>0$ such that, for $k\geq1$ and
$d\geq C_0(1+\log k)$, with probability at least $1-e^{-c_0d}$,
\begin{equation}\label{eq:l2-l1-constant-probability}
    \Ronesum(\alg(R,\locs))
    \leq
    \frac{32e}{3}\beta\Roneopt.
\end{equation}
\end{proposition}

\begin{proof}
Fix $t\geq1$, and consider a rotation $R$ for which
$\frac{\Ronesum(\alg(R,\locs))}{\Roneopt}
    >
    \frac{16}{3}\beta t$.
For this fixed $R$, Claim~\ref{cl:l2-l1-distortion} states that, with
probability at least $1/6$ over the sample $S$,
\[
    \frac{\Ronesum(\alg(R,\locs))}{\Roneopt}
    \leq
    \frac{16}{3}\beta\rho(\mathcal V_S,R).
\]
The left-hand side depends only on $R$ and not on the sampled
instance $S$. Therefore, for every sample on which the conclusion of the
claim holds, the assumed inequality forces
$\rho(\mathcal V_S,R)>t$. It follows that, for every such rotation
$R$, $\prob[S]{\rho(\mathcal V_S,R)>t}
    \geq
    \frac16$.

We now average this statement over $R$. The joint probability can be
written as
\[
    \prob[R,S]{\rho(\mathcal V_S,R)>t}
    =
    \Ex[R]{
        \prob[S]{\rho(\mathcal V_S,R)>t}
    }.
\]
For every rotation satisfying
$\Ronesum(\alg(R,\locs))/\Roneopt>(16/3)\beta t$, the inner probability
is at least $1/6$; for all other rotations it is nonnegative.
Consequently,
\begin{align}
    \frac16
    \prob[R]{
        \frac{\Ronesum(\alg(R,\locs))}{\Roneopt}
        >
        \frac{16}{3}\beta t
    }
    &\leq
    \prob[R,S]{\rho(\mathcal V_S,R)>t}.
    \label{eq:l2-l1-ratio-tail-transfer}
\end{align}

It remains to upper-bound the joint probability on the right. Fix a
sample $S$. Since $\abs{\supp(S)}\leq M$, the family
$\mathcal V_S$ contains at most $M^2$ nonzero pairwise directions.
Moreover, $S$ is sampled independently of $R$, so after conditioning
on $S$, the family $\mathcal V_S$ is fixed before the rotation is
drawn. Corollary~\ref{lem:rotation-family} therefore gives
\[
    \prob[R]{\rho(\mathcal V_S,R)>t\mid S}
    \leq
    \min\left\{
        1,
        M^2\left(\frac et\right)^d
    \right\}.
\]
Averaging this inequality over $S$ yields
\begin{align*}
    \prob[R,S]{\rho(\mathcal V_S,R)>t}=
    \Ex[S]{
        \prob[R]{\rho(\mathcal V_S,R)>t\mid S}
    }\leq
    \min\left\{1,
        M^2\left(\frac et\right)^d
    \right\}.
\end{align*}
Combining this bound with
\eqref{eq:l2-l1-ratio-tail-transfer}, we obtain
\begin{equation}\label{eq:l2-l1-ratio-tail}
    \prob[R]{
        \frac{\Ronesum(\alg(R,X))}{\Roneopt}
        >
        \frac{16}{3}\beta t
    }
    \leq
    \min\left\{
        1,
        6M^2\left(\frac et\right)^d
    \right\}.
\end{equation}

Taking $t=2e$, we obtain
\[
    \prob[R]{
        \frac{\Ronesum(\alg(R,X))}{\Roneopt}
        >
        \frac{32e}{3}\beta
    }
    \leq
    6M^2 2^{-d}.
\]
Because the sample size satisfies
$M\leq C_s k\log^b(ek)$, we have
\begin{align*}
    \log(6M^2)\leq
    \log 6+2\log C_s+2\log k+2b\log\log(ek)\leq
    C_1(1+\log k).
\end{align*}
for an absolute constant $C_1>0$. The last inequality uses
$\log\log(ek)=\log(1+\log k)\leq1+\log k$. Thus the
$\log^b(ek)$ factor in the sample size contributes only
$O(\log\log k)$ after taking logarithms, which is absorbed by
$O(1+\log k)$.

It follows that $6M^2 2^{-d}=
    \exp\left(\log(6M^2)-d\log 2\right)\leq
    \exp\left(C_1(1+\log k)-d\log 2\right)$.
If $d\geq C_0(1+\log k)$, then
$1+\log k\leq d/C_0$, and hence
$6M^2 2^{-d}
    \leq
    \exp\left[
        -\left(\log 2-\frac{C_1}{C_0}\right)d
    \right]$.
Choosing $C_0\geq2C_1/\log 2$ gives $6M^2 2^{-d}
    \leq
    \exp\left(-\frac{\log 2}{2}d\right)
    =
    e^{-c_0d}$,
where $c_0\eqdef(\log 2)/2$.
\end{proof}

We next integrate~\eqref{eq:l2-l1-ratio-tail} to bound the expected
approximation ratio.

\begin{corollary}\label{cor:l2-l1-expectation}
There is an absolute constant $C>0$ such that, for every $k\geq1$
and $d\geq2$,
\begin{equation}\label{eq:l2-l1-expectation}
    \Ex[R]{
        \frac{\Ronesum(\alg(R,\locs))}{\Roneopt}
    }
    \leq
    C\beta\min\{\sqrt d,k^{2/d}\}.
\end{equation}
\end{corollary}

\begin{proof}
Set $T\eqdef e(6M^2)^{1/d}$. Integrating
\eqref{eq:l2-l1-ratio-tail} gives
\begin{align*}
    \Ex[R]{
        \frac{\Ronesum(\alg(R,\locs))}{\Roneopt}
    }
    \leq
    \frac{16\beta}{3}
    \left(
        T+
        6M^2e^d\int_T^\infty t^{-d}\,dt
    \right)=
    \frac{16\beta T}{3}
    \left(1+\frac1{d-1}\right)
    \leq
    C\beta M^{2/d}.
\end{align*}

We also have a pointwise bound. Norm comparison and the
$\beta$-approximation guarantee give
\[
    \Ronesum(\alg(R,\locs))
    \leq
    \sqrt d\,\twosum(\alg(R,\locs))
    \leq
    \beta\sqrt d\,\twoopt.
\]
Since $\Roneopt\geq\twoopt$,
$
    \frac{\Ronesum(\alg(R,\locs))}{\Roneopt}
    \leq
    \beta\sqrt d.
$

It remains to replace $M$ by $k$. If
$k^{2/d}\geq\sqrt d$, the pointwise bound already gives the claim.
Otherwise, $\log k<(d/4)\log d$, and
\begin{align*}
    M^{2/d}
    &\leq
    k^{2/d}
    \exp\left(
        \frac{2\log C_s+2b\log\log(ek)}{d}
    \right)\leq
    Ck^{2/d},
\end{align*}
where the last inequality follows because the exponential factor is
bounded by an absolute constant under
$\log k<(d/4)\log d$. Combining the two cases proves
\eqref{eq:l2-l1-expectation}.
\end{proof}

Optimizing the expectation bound over the dimension gives a guarantee
depending only on $k$.

\begin{corollary}\label{cor:l2-l1-k-only}
For $k\geq3$, uniformly over all $d\geq2$,
\[
    \Ex[R]{
        \frac{\Ronesum(\alg(R,\locs))}{\Roneopt}
    }
    =
    O\left(
        \beta\sqrt{\frac{\log k}{\log\log k}}
    \right).
\]
\end{corollary}

\begin{proof}
By Corollary~\ref{cor:l2-l1-expectation}, it is enough to maximize
$\min\{\sqrt d,k^{2/d}\}$ over $d$. The first term increases with
$d$, while the second decreases. They are equal when
$d\log d=4\log k$, which gives
$d=\Theta(\log k/\log\log k)$. The maximum is therefore
$\Theta(\sqrt{\log k/\log\log k})$.
\end{proof}




\subsection{A lower-bound construction}
\label{subsec:l2-l1-lower}
In this subsection, we show that, even when the returned solution is an
exact $L_2$ optimum for every rotation, the approximation ratio can be
$\Omega(\min\{\sqrt d,k^{1/d}\})$ both with high probability and in
expectation over $R$.
\begin{theorem}\label{thm:l2-l1-lower}
For every $k,d\geq2$, there is an instance with $n=k+1$ points such
that, under $L_1$ objective, the Euclidean optimum has approximation
ratio $\Omega(\min\{\sqrt d,k^{1/d}\})$ with probability at least
$1-(e/4)^d$ over $R$. On the same instance, the expected approximation ratio under the rotated \(L_1\) objective is also
$\Omega(\min\{\sqrt d,k^{1/d}\})$.
\end{theorem}

We reuse the separated-pair construction from
Section~\ref{subsec:l1-l2-lower}. In the present case, the
special pair determines the $L_1$ cost of the Euclidean solution,
while a nearly axis-aligned candidate pair gives an upper bound on
$\Roneopt$.

Let
$T=\frac1{25}\min\{\sqrt d,k^{1/d}\}$ and
$\eps=1/(5T)$. When $T>2$, choose an $\eps$-net
$U=\{u_1,\ldots,u_s\}$ as in the proof for
Theorem~\ref{thm:l1-l2-lower}; the same calculation gives
$s\leq(k-1)/2$. Place the special pair at $-e_1/2$ and $e_1/2$.
For each $u_r\in U$, place a candidate pair at
$b_r-u_r$ and $b_r+u_r$, where
$b_r\eqdef1000r\sqrt d\,e_1$. Place the remaining points as
singletons at additional sites $b_r$. The sites are far enough apart
that each designated pair is isolated from all other points.

\begin{claim*}
On the instance described above, the Euclidean
optimum $\Ctwo$ satisfies
\begin{align}
    \prob[R]{
        \frac{\Ronesum(\Ctwo)}{\Roneopt}
        \geq
        \frac{\min\{\sqrt d,k^{1/d}\}}{200}
    }
    &\geq
    1-\left(\frac e4\right)^d \label{eq:l2-l1-lower-probability}\\
    \Ex[R]{
        \frac{\Ronesum(\Ctwo)}{\Roneopt}
    }
    &\geq
    \frac{\min\{\sqrt d,k^{1/d}\}}{200}.
    \label{eq:l2-l1-lower-ratio-of-expectations}
\end{align}

\end{claim*}

\begin{proof}
If $T\leq2$, take any instance with $k+1$ distinct points and positive
optimal cost. Both ratios are at least one,
so the conclusions hold with the stated constant $1/200$. We therefore
assume $T>2$ and use the construction above.

The special pair is the unique Euclidean closest pair. Choose
$\Ctwo$ to contain $e_1/2$ and every point outside the special pair.
Then $\twoopt=1$, and the only point with positive cost is
$-e_1/2$. Since all other sites are remote,
$\Ronesum(\Ctwo)=\onenorm{Re_1}$, and hence
$\Ex[R]{\Ronesum(\Ctwo)}=\lambda_d$.

For every rotation $R$, the net contains some $u_r$ satisfying
$\twonorm{Ru_r-e_1}\leq1/(5T)$. Serving the candidate pair
$b_r-u_r,b_r+u_r$ together and placing centers at all remaining
points gives
\[
    \Roneopt
    \leq
    2\onenorm{Ru_r}
    \leq
    2+\frac{2\sqrt d}{5T}
    \leq
    \frac{\sqrt d}{T},
\]
where the last inequality uses $T\leq\sqrt d/25$.

Consider the event
$\mathcal A\eqdef\{\onenorm{Re_1}\geq\sqrt d/4\}$.
Proposition~\ref{lem:rotation-small-ball}, applied to $e_1$ with
$t=4$, gives
$\prob[R]{\mathcal A}\geq1-(e/4)^d$.
In the event $\mathcal A$,
\[
    \frac{\Ronesum(\Ctwo)}{\Roneopt}
    \geq
    \frac{\sqrt d/4}{\sqrt d/T}
    =
    \frac T4
    =
    \frac1{100}\min\{\sqrt d,k^{1/d}\}.
\]
This proves~\eqref{eq:l2-l1-lower-probability} with the stated
constant $1/200$. Since
$1-(e/4)^2>1/2$,
\[
    \Ex[R]{
        \frac{\Ronesum(\Ctwo)}{\Roneopt}
    }
    \geq
    \frac T4\prob[R]{\mathcal A}
    \geq
    \frac T8
    =
    \frac1{200}\min\{\sqrt d,k^{1/d}\}.
\]
This proves both conclusions.
\end{proof}
Further maximizing over $d$ yields the following asymptotically tight bound on $k$.
\begin{corollary}\label{cor:l2-l1-tight-k-dependence}
For every $k\geq3$, the worst-case expected approximation ratio of an
exact $L_2$ optimum under the rotated $L_1$ objective is
$\Theta\left(
        \sqrt{\frac{\log k}{\log\log k}}
    \right)$.
\end{corollary}

\begin{proof}
The upper bound follows from
Corollary~\ref{cor:l2-l1-k-only} with $\beta=1$.
For the lower bound, Theorem~\ref{thm:l2-l1-lower} gives
$\Omega(\min\{\sqrt d,k^{1/d}\})$. Optimizing this expression over
$d$ exactly as in the proof of
Corollary~\ref{cor:l1-l2-tight-k-dependence} gives
$\Omega(\sqrt{\log k/\log\log k})$.
\end{proof}

\section{Discussion}
We study the mismatch between the Manhattan and Euclidean
$k$-median objectives when the input instance is arbitrary but the
coordinate system is chosen uniformly at random. In both directions,
the main difficulty is that the clustering returned by the algorithm
may depend on the rotation. Our proofs address this dependence by
reducing the analysis to a small family of directions fixed
independently of the rotation: differences between the centers of an
optimal Euclidean solution in one direction, and differences between
points in a small weighted coreset in the other.

These reductions give constant-factor guarantees with high
probability when $d=\Omega(\log k)$ and expected approximation ratios
of
$O(\beta\min\{\sqrt d,k^{2/d}\})$ in both directions. The lower-bound
constructions show that even exact optimal solutions can have expected
ratio
$\Omega(\min\{\sqrt d,k^{1/d}\})$. Although a gap remains for fixed
$k$ and $d$, the bounds have the same worst-case dependence
$\Theta(\sqrt{\log k/\log\log k})$ after maximizing over the
dimension. Thus, random rotation substantially reduces the
worst-case $\sqrt d$ mismatch, but it does not remove the difference
between the two objectives completely.

A first open question is to close the gap between $k^{1/d}$ and
$k^{2/d}$ for a fixed dimension. This requires either a sharper
analysis of the finite families of directions arising from projection
and sampling, or a lower-bound construction with a larger family of
simultaneously unfavorable directions.

Another direction is to return to the comparison that motivates the
introduction: the $L_1$ objective for $k$-median and the squared
Euclidean objective $L_2^2$ for $k$-means. Our present analysis uses
the unsquared Euclidean objective in order to isolate the effect of changing the ambient norm from the separate effect of squaring
distances. The one-dimensional example in the introduction shows that
the latter effect can already produce a large mismatch when the
$L_1$ and $L_2$ norms coincide. Therefore, randomizing the coordinate
system alone cannot guarantee that $L_1$ and $L_2^2$ solutions are
close on every instance. The natural question is instead to characterize the instances for
which the two objectives have simultaneously good solutions. For
example, one may ask whether bounded tails, balanced cluster sizes, or
a controlled relation between the first and second moments of the
within-cluster distances ensures that an approximate solution for one
objective is also a good approximation for the other. 
\appendix
\section{Deferred proofs}
    \label{sec:appendix}
\subsection{Proof of Claim~\ref{cl:single_cluster_l2_under_l1}}
\label{app:one-center-l2-under-l1}
We first prove the upper bound. If $g=m$, then
$\onesum(g)=\oneopt$. Otherwise, translate the instance so that $g=0$
and reflect individual coordinate axes so that $m_\ell\geq0$ for every
$\ell\in[d]$. These transformations preserve both objectives and the
optimality of the two medians. Hence,
$\twoopt=\sum_{i=1}^n\twonorm{\loci}$, and $\oneopt=\sum_{i=1}^n\onenorm{\loci-m}$.
By the triangle inequality,
\begin{equation}\label{eq:oc-displacement-reduction}
    \onesum(g)
    \leq
    \onesum(m)+n\onenorm{g-m}
    =
    \oneopt+n\onenorm{m}.
\end{equation}
It therefore suffices to bound the distance $\onenorm{m}$ between the
two medians.
For every point $\loci\neq0$, define $u_i\eqdef\frac{\loci}{\twonorm{\loci}}$.
The vector $u_i$ is the unit directional contribution of point $i$ at
the geometric median. Our proof compares the positive and negative
coordinate contributions of these vectors. The coordinate-median
condition will force a large positive contribution, while
geometric-median optimality requires this contribution to be balanced
by an equal negative contribution. We will then use the Manhattan cost
at $m$ to upper-bound how large the negative contribution can be.

We begin by recording the balance condition implied by the optimality
of the geometric median. For points located at the origin, the
corresponding directional vectors are not uniquely determined, but they
can be chosen from the Euclidean unit ball so that all directional
contributions balance.

\begin{lemma}
\label{lem:opt_condition}
For every $i$ with $\loci=0$, one can choose a vector $u_i\in\R^d$
such that
\begin{equation}\label{eq:oc-force-balance}
    \twonorm{u_i}\leq1
    \quad\text{for every }i\in[n],
    \qquad
    \sum_{i=1}^n u_i=0.
\end{equation}
For every $i$ with $\loci\neq0$, we retain the definition
$u_i=\loci/\twonorm{\loci}$.
\end{lemma}
\begin{proof}
    Let $q$ be the number of points at the origin. For every
Euclidean unit vector $v$, optimality of $0$ gives
$0\leq
    \left.\frac{d}{dt}\twosum(tv)\right|_{t=0^+}
    =q-\sum_{i:\loci\neq0}\iprod{u_i}{v}$.
It follows that $\twonorm{\sum_{i:\loci\neq0}u_i}\leq q$.
If $q>0$, assign $-\sum_{i:\loci\neq0}u_i/q$ to each origin point. If $q=0$, the sum of the nonzero points' vectors is already zero.
This proves~\eqref{eq:oc-force-balance}.
\end{proof}
Write $u_{i\ell}$ for the $\ell$th coordinate of $u_i$.
For a scalar $t$, write $t_+\eqdef\max\{t,0\}$ and
$t_-\eqdef\max\{-t,0\}$. Equation~\eqref{eq:oc-force-balance}
implies that the positive and negative contributions agree in each
coordinate, and therefore
\begin{equation}\label{eq:oc-force-total}
    \sum_{i=1}^n\sum_{\ell=1}^d(u_{i\ell})_+
    =
    \sum_{i=1}^n\sum_{\ell=1}^d(u_{i\ell})_-.
\end{equation}
We first lower-bound the positive side. When $m_\ell>0$, at least half
the points have $(\loci)_\ell\geq m_\ell$ and contribute positively in coordinate
$\ell$. Moving these points farther away can reduce their unit-vector
components, but their total Euclidean distance is bounded by $\twoopt$.

\begin{lemma}\label{lem:oc-force-lower}
    $\sum_{i=1}^n\sum_{\ell=1}^d(u_{i\ell})_+
    \geq\frac{n^2\onenorm{m}}{4\twoopt}$.
\end{lemma}

\begin{proof}
Fix a coordinate $\ell$ with $m_\ell>0$ and let $A_\ell\eqdef\{i:(\loci)_\ell\geq m_\ell\}$.
The coordinate-wise median condition gives $\abs{A_\ell}\geq n/2$. For every $i\in A_\ell$,
we have $\loci\neq0$ and
$u_{i\ell}=\frac{(\loci)_\ell}{\twonorm{\loci}}
    \geq\frac{m_\ell}{\twonorm{\loci}}>0$.
Consequently,
\begin{equation}\label{eq:oc-positive-coordinate}
    \sum_{i=1}^n(u_{i\ell})_+
    \geq m_\ell\sum_{i\in A_\ell}\frac1{\twonorm{\loci}}.
\end{equation}
By Cauchy--Schwarz,
\[
    \frac{n^2}{4}
    \leq\abs{A_\ell}^2
    \leq
    \left(\sum_{i\in A_\ell}\twonorm{\loci}\right)
    \left(\sum_{i\in A_\ell}\frac1{\twonorm{\loci}}\right)
    \leq
    \twoopt\sum_{i\in A_\ell}\frac1{\twonorm{\loci}}.
\]
Substituting into~\eqref{eq:oc-positive-coordinate} yields
$\sum_{i=1}^n(u_{i\ell})_+
    \geq\frac{n^2m_\ell}{4\twoopt}$.
Summing over the coordinates with $m_\ell>0$, we obtain
\[
    \sum_{i=1}^n\sum_{\ell=1}^d(u_{i\ell})_+
    \geq
    \sum_{\ell:m_\ell>0}\frac{n^2m_\ell}{4\twoopt}
    =\frac{n^2\onenorm{m}}{4\twoopt}.
    \qedhere
\]
\end{proof}

We next upper-bound the negative side of~\eqref{eq:oc-force-total}
using $\oneopt$. If $\loci\neq0$ and $u_{i\ell}<0$, then
$(\loci)_\ell<0\leq m_\ell$, so 
\begin{align}
    \label{eq:oc-negative-charge}
    \abs{(\loci)_\ell-m_\ell}
    \geq\abs{(\loci)_\ell}
    =\twonorm{\loci}(u_{i\ell})_-.
\end{align} 
When $\twonorm{\loci}$ is large, then the inequality already provides a meaningful lower bound for $\abs{(\loci)_\ell-m_\ell}$. On the other hand, when $\twonorm{\loci}$ is small, we show that the number of such points can be estimated.

\begin{lemma}\label{lem:oc-force-upper}
For every
$0<r<\onenorm{m}/\sqrt d$,
\begin{equation}\label{eq:oc-force-radius}
    \sum_{i=1}^n\sum_{\ell=1}^d(u_{i\ell})_-
    \leq
    \frac{\sqrt d\,\oneopt}{\onenorm{m}-\sqrt d\,r}
    +\frac{\oneopt}{r}.
\end{equation}
In particular,
\begin{equation}\label{eq:oc-force-upper}
    \sum_{i=1}^n\sum_{\ell=1}^d(u_{i\ell})_-
    \leq\frac{4\sqrt d\,\oneopt}{\onenorm{m}}.
\end{equation}
\end{lemma}

\begin{proof}
We split at a radius $r$ and then choose $r$
to balance the two bounds. If $\twonorm{\loci}\leq r$, then
$\onenorm{\loci-m}
    \geq\onenorm{m}-\onenorm{\loci}
    \geq\onenorm{m}-\sqrt d\,r$.
There are therefore at most $\oneopt/(\onenorm{m}-\sqrt d\,r)$
such points. Each contributes at most $\sum_{\ell=1}^d(u_{i\ell})_-
    \leq\onenorm{u_i}
    \leq\sqrt d\,\twonorm{u_i}
    \leq\sqrt d$.
Hence
\begin{equation}\label{eq:oc-near-force}
    \sum_{i:\twonorm{\loci}\leq r}\sum_{\ell=1}^d(u_{i\ell})_-
    \leq\frac{\sqrt d\,\oneopt}{\onenorm{m}-\sqrt d\,r}.
\end{equation}
For the remaining points, equation~\eqref{eq:oc-negative-charge} gives
\begin{equation}\label{eq:oc-far-force}
    \sum_{i:\twonorm{\loci}>r}\sum_{\ell=1}^d(u_{i\ell})_-
    \leq\frac1r\sum_{i:\twonorm{\loci}>r}\onenorm{\loci-m}
    \leq\frac{\oneopt}{r}.
\end{equation}
Adding~\eqref{eq:oc-near-force} and~\eqref{eq:oc-far-force}
proves~\eqref{eq:oc-force-radius}. Its right-hand side is
$\frac{\onenorm{m}\,\oneopt}
         {r(\onenorm{m}-\sqrt d\,r)}$.
The denominator is maximized at $r=\onenorm{m}/(2\sqrt d)$.
At this radius the two bounds are equal, and their sum is
$4\sqrt d\,\oneopt/\onenorm{m}$, proving~\eqref{eq:oc-force-upper}.
\end{proof}
Combining Lemmas~\ref{lem:oc-force-lower} with~\ref{lem:oc-force-upper},
and equation~\eqref{eq:oc-force-total} gives
$\frac{n^2\onenorm{m}}{4\twoopt}
    \leq\sum_{i,\ell}(u_{i\ell})_+
    =\sum_{i,\ell}(u_{i\ell})_-
    \leq\frac{4\sqrt d\,\oneopt}{\onenorm{m}}$.
Multiplying through yields
$n^2\onenorm{m}^2\leq16\sqrt d\,\twoopt\oneopt$.
Taking square roots and returning to the original coordinates,
\begin{equation}\label{eq:oc-displacement}
    n\onenorm{g-m}
    \leq4d^{1/4}\sqrt{\twoopt\oneopt}.
\end{equation}
Since $\twoopt=\twosum(g)\leq \twosum(m)\leq \onesum(m)=\oneopt$,
equation~\eqref{eq:oc-displacement-reduction} now gives
\[
    \onesum(g)
    \leq\oneopt+4d^{1/4}\sqrt{\twoopt\oneopt}
    \leq\bigl(1+4d^{1/4}\bigr)\oneopt.
    \qedhere
\]
We next derive the matching lower bound. To obtain the same dependence on $d$, we keep the geometric median at
the origin while shifting the coordinate-wise median in every
coordinate. Each of $d$ points has one large coordinate and the same
small value in all other coordinates. Moving the Manhattan center to
that common offset removes all the small-coordinate costs at once.
Additional points at the origin keep the geometric median there.
We choose their number below $d$ so that they do not determine the
coordinate-wise median.

\begin{lemma}\label{lem:oc-lower}
For every $d\geq1$, there is a location profile with a unique geometric
median $g$ and a unique coordinate-wise median $m$ such that
\begin{equation}\label{eq:oc-lower}
    \frac{\onesum(g)}{\oneopt}\geq\frac18d^{1/4}.
\end{equation}
Consequently, the worst-case ratio for an $L_2$ optimum under the $L_1$ objective is
$\Theta(d^{1/4})$.
\end{lemma}

\begin{proof}
First suppose $d\geq256$. Let $e_1,\ldots,e_d$ be the standard basis
vectors and let $\mathbf{1}=(1,\ldots,1)$. Set
   $b= d^{-3/4}$ and $h=\sqrt{1-(d-1)b^2}$.
Place one point at each of the $d$ locations
    $\loci= h e_i+b\sum_{\ell\neq i}e_\ell$, for $1\leq i\leq d$,
and place another
    $q=\left\lceil2\sqrt d\,\bigl(h+(d-1)b\bigr)\right\rceil$
points at the origin. Thus $n=d+q$ and $\loci[d+1]=\cdots=\loci[d+q]=0$.
The choice of $h$ makes each nonzero point a Euclidean unit vector, with $\onenorm{\loci}=h+(d-1)b$ and
    $\sum_{i=1}^d \loci=\bigl(h+(d-1)b\bigr)\mathbf{1}$.

We first verify that the origin is the unique geometric median.
For every $z\in\R^d$ and each nonzero $\loci$, Cauchy--Schwarz gives $\twonorm{\loci-z}=\twonorm{\loci-z}\cdot \twonorm{\loci}
    \geq\iprod{\loci}{\loci-z}
    =1-\iprod{\loci}{z}$.
Summing over the $d$ nonzero points and adding the origin points,
\begin{align*}
    \twosum(z)
    \geq d+q\twonorm{z}
       -\bigl(h+(d-1)b\bigr)\iprod{\mathbf{1}}{z}
    \geq \twosum(0)
       +\bigl(q-\sqrt d\,(h+(d-1)b)\bigr)\twonorm{z}.
\end{align*}
The coefficient of $\twonorm{z}$ is positive, so $g=0$ is the unique
geometric median.

Next, we identify the coordinate-wise median. We have $h>b>0$, since
$h^2-b^2=1-d^{-1/2}>0$. In each coordinate, the values are $q$ copies
of $0$, $d-1$ copies of $b$, and one copy of $h$. Since $h\leq1$ and
$(d-1)b\leq d^{1/4}$, $\frac qd
    \leq2d^{-1/2}+2d^{-1/4}+d^{-1}<1$ when $d\geq256$.
Thus $q<d$, so fewer than half the points lie strictly below $b$;
only one lies strictly above it. The unique median in every
coordinate is therefore $b$, and $m=b\mathbf{1}$.

It remains to compare the Manhattan costs at these two centers.
At the origin, each nonzero point contributes $h+(d-1)b$, giving $\onesum(g)=\onesum(0)=d\bigl(h+(d-1)b\bigr)\geq d(d-1)b$.
At $m=b\mathbf{1}$, each nonzero point differs from $m$ only in its
large coordinate and contributes $h-b$. Each origin point contributes
$db$. Hence $\oneopt=\onesum(m)=d(h-b)+qdb=d(h-b+qb)$, where
\begin{align*}
    h-b+qb\leq1+qb\leq1+\bigl(2\sqrt d\,(1+d^{1/4})+1\bigr)d^{-3/4}=3+2d^{-1/4}+d^{-3/4}<4.
\end{align*}
Consequently,
\[
    \frac{\onesum(g)}{\oneopt}
    =\frac{h+(d-1)b}{h-b+qb}\geq \frac{(d-1)b}{4}
    \geq\frac{d^{1/4}-d^{-3/4}}{4}
    \geq\frac18d^{1/4}.
\]

For $1\leq d<256$, take three points at $-e_1$, $0$, and $e_1$.
Both medians are uniquely $0$, and the ratio is $1$, which also
satisfies~\eqref{eq:oc-lower}.
\end{proof}

\subsection{Proofs of the rotation estimates}
\label{app:rotation-estimates}

We give the proofs of the estimates stated in
Section~\ref{sec:preliminary}. The expected-length estimate
uses the Gaussian representation of spherical measure, the small-ball
estimate follows from a cone-volume comparison, and the final estimate
combines spherical concentration with convexity.

\begin{proof}[Proof of Proposition~\ref{lem:rotation-mean}]
For $v\neq0$, the vector $Rv/\twonorm{v}$ is uniformly distributed on
$\mathbb S^{d-1}$. Homogeneity therefore gives
$\Ex[R]{\onenorm{Rv}}
    =
    \twonorm{v}\,
    \Ex[R]{\onenorm{Rv/\twonorm{v}}}
    =
    \lambda_d\twonorm{v}$.
The identity is immediate for $v=0$. Also,
$\onenorm{V}\leq\sqrt d\,\twonorm{V}=\sqrt d$, so
$\lambda_d\leq\sqrt d$.

For the lower bound, let $G=(G_1,\ldots,G_d)$ have independent
standard normal coordinates, and define
$T\eqdef\twonorm{G},
    U\eqdef G/T$.
The event $T=0$ has probability zero. The Gaussian representation of
spherical measure states that $U$ is uniform on $\mathbb S^{d-1}$
and independent of $T$; see
\cite[Exercise~3.3.7]{Vershynin_2018}. Consequently,
$\Ex{\onenorm{G}}=\Ex{T\onenorm{U}}=(\Ex{T})\lambda_d$.
Each coordinate satisfies $\Ex{\abs{G_\ell}}
    =
    \frac{2}{\sqrt{2\pi}}
    \int_0^\infty t e^{-t^2/2}\,dt
    =
    \sqrt{\frac2\pi}$,
and hence $\Ex{\onenorm{G}}=d\sqrt{2/\pi}$.
By Cauchy--Schwarz,
$\Ex{T}
    \leq
    \sqrt{\Ex{T^2}}
    =
    \sqrt{\sum_{\ell=1}^d\Ex{G_\ell^2}}
    =
    \sqrt d$.
It follows that $ d\sqrt{\frac2\pi}
    =
    (\Ex{T})\lambda_d
    \leq
    \sqrt d\,\lambda_d$,
which proves the lower bound.
\end{proof}

\begin{proof}[Proof of Proposition~\ref{lem:rotation-small-ball}]
We specialize the standard polar-coordinate volume comparison (see~\cite[Lecture~1]{Ball1997}).
Fix $v\neq0$ and put $V\eqdef\frac{Rv}{\twonorm{v}}$.
Then $V$ is uniform on $\mathbb S^{d-1}$. Define
$A_t\eqdef\left\{u\in\mathbb S^{d-1}:\onenorm{u}\leq\frac{\sqrt d}{t}\right\}$,
the set of unit directions whose Manhattan length is at most
$\sqrt d/t$. By homogeneity, $\prob[R]{\onenorm{Rv}
        \leq\frac{\sqrt d}{t}\twonorm{v}}=\prob{V\in A_t}$.
Let $B_1^d\eqdef\{x\in\R^d:\onenorm{x}\leq1\},
    \qquad
    B_2^d\eqdef\{x\in\R^d:\twonorm{x}\leq1\}$
be the $L_1$ and Euclidean unit balls. For $a>0$, the scaled ball
$aB_1^d=\{ax:x\in B_1^d\}$ consists of the points with Manhattan
norm at most $a$.

To turn the spherical probability into a volume ratio, define $D_t
    \eqdef
    \{ru:u\in A_t,\ 0\leq r\leq1\}$.
This is the union of the line segments joining the origin to the
directions in $A_t$. Write $\sigma$ for surface-area measure on the
sphere and $\operatorname{vol}$ for $d$-dimensional volume.
Polar coordinates give
$\operatorname{vol}(D_t)=
    \int_{A_t}\int_0^1 r^{d-1}\,dr\,d\sigma(u) =
    \frac{\sigma(A_t)}d$, and
    $\operatorname{vol}(B_2^d)
    =
    \frac{\sigma(\mathbb S^{d-1})}d$.
Thus
$\prob{V\in A_t}=
    \frac{\sigma(A_t)}{\sigma(\mathbb S^{d-1})}=
    \frac{\operatorname{vol}(D_t)}{\operatorname{vol}(B_2^d)}$.

Every $x=ru\in D_t$ satisfies
$\onenorm{x}=r\onenorm{u}\leq\sqrt d/t$, so
$D_t\subseteq(\sqrt d/t)B_1^d$. Volume scales by the $d$th power of
the scaling factor, and therefore
\begin{align}
    \prob{V\in A_t}=
    \frac{\operatorname{vol}(D_t)}
         {\operatorname{vol}(B_2^d)}\leq
    \left(\frac{\sqrt d}{t}\right)^d
    \frac{\operatorname{vol}(B_1^d)}
         {\operatorname{vol}(B_2^d)}.
    \label{eq:rotation-volume}
\end{align}

The $\ell_1$ unit ball is the union of $2^d$ congruent simplices,
one in each coordinate orthant. Each has volume $1/d!$, so
$\operatorname{vol}(B_1^d)=\frac{2^d}{d!}$.
The cube $[-1/\sqrt d,1/\sqrt d]^d$ lies in $B_2^d$, since every
point in the cube has squared Euclidean norm at most
$d(1/d)=1$. Hence $\operatorname{vol}(B_2^d)
    \geq
    \left(\frac2{\sqrt d}\right)^d$.
Substitution into~\eqref{eq:rotation-volume} gives
$\prob{V\in A_t}
    \leq
    \left(\frac{\sqrt d}{t}\right)^d
    \frac{2^d/d!}{(2/\sqrt d)^d}
    =
    \frac{d^d}{d!\,t^d}$.
Finally, $\log(d!)
    =
    \sum_{\ell=1}^d\log\ell
    \geq
    \int_1^d\log x\,dx
    \geq
    d\log d-d$,
so $d!\geq(d/e)^d$. We obtain
$\prob{V\in A_t}\leq(e/t)^d$. Combining this with the upper bound
one on every probability proves the proposition.
\end{proof}

\begin{proof}[Proof of Proposition~\ref{lem:rotation-convex-concentration}]
We first apply spherical concentration to the length of a single
unit vector. A function $f:\mathbb S^{d-1}\to\R$ is $L$-Lipschitz
if $\abs{f(u)-f(v)}
    \leq L\twonorm{u-v}$
 for all $u,v\in\mathbb S^{d-1}$.
For a uniform sphere point $V$, concentration on the unit sphere
gives
\[
    \prob{
        \abs{f(V)-\Ex{f(V)}}>t
    }
    \leq
    2\exp\left(-\frac{c d t^2}{L^2}\right),
    \qquad t>0,
\]
where $c>0$ is absolute; see
\cite[Theorem~5.1.4]{Vershynin_2018}.

Since $\abs{\onenorm{u}-\onenorm{v}}
    \leq
    \onenorm{u-v}
    \leq
    \sqrt d\,\twonorm{u-v}$,
the function $u\mapsto\onenorm{u}$ is $\sqrt d$-Lipschitz.
Therefore, for $X=\onenorm{V}-\lambda_d$,
$\prob{\abs X>t}\leq2e^{-ct^2}$.
Integrating this tail gives an exponential-moment bound:
\begin{align*}
    \Ex{\exp(cX^2/4)}
    =
    1+\int_1^\infty
        \prob{\exp(cX^2/4)>s}\,ds\leq
    1+2\int_1^\infty s^{-4}\,ds
    =
    \frac53
    <2.
\end{align*}

Return to the vectors and coefficients in the proposition. Omit
terms with $b_j\twonorm{v_j}=0$, and normalize
    $u_j\eqdef\frac{v_j}{\twonorm{v_j}}$,
    $\beta_j\eqdef\frac{b_j\twonorm{v_j}}D$, and then write
    $X_j(R)\eqdef\onenorm{Ru_j}-\lambda_d$.
Thus $\beta_j\geq0$, $\sum_j\beta_j=1$, and
$\frac1D\sum_j b_j\onenorm{Rv_j}-\lambda_d=
    \sum_j\beta_jX_j(R)$.
Each $Ru_j$ is uniform on the sphere, so the exponential-moment
bound above applies to every $X_j(R)$. The function
$x\mapsto\exp(cx^2/4)$ is convex. Jensen's inequality therefore gives
\[
    \Ex[R]{
        \exp\left(
            \frac c4\left(\sum_j\beta_jX_j(R)\right)^2
        \right)
    }
    \leq
    \sum_j\beta_j
        \Ex[R]{\exp(cX_j(R)^2/4)}
    \leq2.
\]
The same rotation is used in all summands; this step uses convexity
and their individual distributions.

Applying Markov's inequality to this exponential, we obtain
\begin{align*}
    \prob[R]{
        \abs{\frac1D\sum_jb_j\onenorm{Rv_j}-\lambda_d}
        >
        \varepsilon\lambda_d
    }
    \leq
    2\exp\left(-\frac c4\varepsilon^2\lambda_d^2\right)
    \leq
    2\exp\left(-\frac{c}{2\pi}\varepsilon^2d\right),
\end{align*}
where the last inequality uses
Proposition~\ref{lem:rotation-mean}. Taking $\kappa=c/(2\pi)$
proves the claim.
\end{proof}

\subsection{Proof of Inequality~\eqref{eq:l2-l1-sampled-optimum-expectation}}
\label{app:coreset_property}
\begin{proof}
   Since $\CRone$ is feasible for every sample,
$L_{1,R}(\OPT_1(R,S);S)\leq L_{1,R}(\CRone;S)$, which gives $\Ex[S]{L_{1,R}(\OPT_1(R,S);S)}\leq\Ex[S]{L_{1,R}(\CRone;S)}$. It remains to show that
$$\Ex[S]{L_{1,R}(\CRone;S)}\leq \Roneopt.$$ 
To prove this, we use the unbiasedness of the sampling: the method assigns a positive sampling probability $\pi_i$ to every
point $\loci$. It then draws $J_1,\ldots,J_M$ independently, where
$\prob{J_t=i}=\pi_i$, and gives the $t$-th sampled point weight
$1/(M\pi_{J_t})$. Define
$d_{1,R}(i)\eqdef\min_{c\in\CRone}\onenorm{R(\loci-c)}$.
In this way, $L_{1,R}(\CRone;S)$ can be written as
$\frac1M\sum_{t=1}^M\frac{d_{1,R}(J_t)}{\pi_{J_t}}$.
Since $\prob{J_t=i}=\pi_i$,
\[
    \Ex[S]{L_{1,R}(\CRone;S)}=
    \frac1M\sum_{t=1}^M\sum_{i=1}^n
        \pi_i\frac{d_{1,R}(i)}{\pi_i}=
    \sum_{i=1}^n d_{1,R}(i)=
    \Roneopt.
\]
\end{proof}

\section*{AI Disclosure}
The authors used ChatGPT 5.6 to assist with discovering relevant background results on
high-dimensional geometry and random rotations. The tool was also
used as an editorial aid to review the exposition and suggest
simplifications and streamlining of arguments originally developed
by the authors. The authors independently verified all mathematical
claims, proofs, and references and take full responsibility for the
contents of the paper.

\bibliographystyle{alpha}
\bibliography{facility}

\end{document}